\documentclass[review]{elsarticle}

\usepackage{lineno,hyperref}
\modulolinenumbers[5]
\usepackage[all]{hypcap}

\usepackage{amssymb}
\usepackage[fleqn]{amsmath}		
\usepackage{graphicx}
\usepackage{comment}

\newcommand*\diff{\mathop{}\!\mathrm{d}}

\usepackage{color,soul}

\usepackage[greek,english]{babel}
\newcommand{\gmu}{\greektext m\latintext}

\usepackage[normalem]{ulem}

\biboptions{numbers,sort&compress}

\usepackage[utf8]{inputenc}
\usepackage{cleveref}
\crefname{section}{§}{§§}
\Crefname{section}{§}{§§}

\journal{International Journal of Hydrogen Energy}

\begin{document}

\begin{frontmatter}

\title{Experimental and Numerical Investigation of Turbulent Jets Issuing Through a Realistic Pipeline Geometry:  Asymmetry Effects for Air, Helium, and Hydrogen}

\author[uvic]{Majid Soleimani nia\corref{cor1}}
\ead{majids@uvic.ca}
\author[uvic]{Brian Maxwell}
\author[uvic]{Peter Oshkai}
\author[uvic]{Nedjib Djilali}

\cortext[cor1]{Corresponding Author}
\address[uvic]{Institute for Integrated Energy Systems, University of Victoria, PO Box 1700 STN CSC, Victoria, Canada, V8W 2Y2}

\begin{abstract}
 Experiments and numerical simulations were conducted to investigate the dispersion of turbulent jets issuing from realistic pipe geometries.  The  effect of jet densities and Reynolds numbers on vertical buoyant  jets were investigated, as they emerged from the side wall of a circular pipe, through a round orifice.  Particle image velocimetry (PIV) and planar laser-induced fluorescence (PLIF) techniques were employed simultaneously to provide time-averaged flow velocity and concentrations fields.  Large eddy simulation (LES) was applied to provide further detail with regards to the three-dimensionality of air, helium, and hydrogen jets.  These realistic jets were always asymmetric and found to deflect about the vertical axis.  This deflection was influenced by buoyancy, where heavier gases deflected more than lighter gases. Significant turbulent mixing was also observed in the near field.  The realistic jets, therefore, experienced faster velocity decay, and asymmetric jet spreading compared to round jets.  These findings indicate that conventional round jet assumptions are, to some extent, inadequate to predict gas concentration, entrainment rates and, consequently, the extent of the flammability envelope of realistic gas leaks.
\end{abstract}


\end{frontmatter}



\section{Introduction}

	Worldwide efforts continue to improve renewable energy technologies, as alternatives for traditional power supply in the energy grid and transportation applications. Hydrogen, as one renewable energy source, can burn or react with almost no pollution.  Commonly, it is used in electrochemical fuel cells to power vehicle and electrical devices.  It can also be burned directly in engines. However, modern safety standards for hydrogen infrastructure must be assured before widespread public use can become possible.  As a result, there has been much focus on advancing research to understand ignition behaviour of hydrogen leaks in order to assess {associated} safety hazards.  {To date, a} number of experiments have shown that hydrogen jets are easily ignitable \cite{Veser2011IJoHE2351}, and have a wide range of ignition limits (between 4\% to 75\% by volume) \cite{Lewis1961}.  It is therefore of paramount interest to understand the dispersive nature of hydrogen, {which is} a highly compressible gas, in order to adequately develop codes and standards. The current study addressed, through experimental measurements and numerical simulation, the effect of jet exit conditions on the dispersion of fuel leaks from a realistic {pipeline} geometry.  The piping arrangement considered here was novel, as we examined the dispersion of vertical jets which emerged through a circular hole located in the side wall of a round pipe, perpendicular to the mean flow within the pipe.  The aim was to {provide insight into} the flow structures associated with hydrogen outflow from a realistic fuel leak scenario.

	Traditionally, scientific research has been limited to compressible fuel leaks through flat surfaces, aligned in the direction of the mean flow origin.   To date, much is known about the axisymmetric and self-similar nature of such jet configurations, emerging through round holes, for a wide range of Reynolds numbers and gas densities.   A recent review on round turbulent jets \cite{Ball2012PiAS1} presented experimental and numerical advances for {a} period of 86 years, starting with the work of Tollmien (1926) \cite{Tollmien1926Z-JoAMaM/ZfAMuM468}.  Of these, significant advances in round jet theory have been made possible from statistical analysis of many physical experiments \cite{Donaldson1966AJ2017,Kapner1970IJHMT932,Grandmaison1982CJCE212,So1990EiF273,Pitts1991EiF125,Pitts1991EiF135,Panchapakesan1993JoFM197,Panchapakesan1993JoFM225,Hussein1994JoFM31,Amielh1996IJoHaMT2149,Djeridane1996PF1614,L.K.2010JoFM59,DeGregorio2014EiF1693}.  Advances in computational resources have also allowed numerical simulation, through large eddy simulation (LES), to prove useful for determining entire flow fields of such round jets \cite{DeBonis2002AJ1346,Suto2004HTAR175,Bogey201013,Chernyavsky2011IJoHE2645,Tajallipour2013IJNMH336}.  In most experiments, data has been collected for air, helium, and $\textrm{CO}_2$ jets, due to the reactive nature of hydrogen.  However, numerical simulation have also proved useful for determining ignition limits associated with hydrogen \cite{Chernyavsky2011IJoHE2645}.  In general, one can categorize a round jet nozzle type through a flat surface as a sharp-edged orifice plate (OP), smooth contraction (SC), or a long pipe (LP). Among these three different nozzles, the most detailed research was performed on SC nozzles \cite{Wygnanski1969JoFM577,Panchapakesan1993JoFM225}.  It has been shown that SC jets have a nearly laminar flow profile at the jet exit with a uniform `top-hat' velocity profile. LP nozzles \cite{Pitts1991EiF125,Pitts1991EiF135,J.2001JoFM91}, on the other hand, produce a nearly Gaussian velocity profile due to fully developed turbulent conditions at the pipe exit.  These jets also have thicker initial shear layers compared to the SC jets. Sharp-edged OP jets have received more recent attention, in the last decade, where detailed measurements \cite{Mi2007EiF625,Quinn2006EJoM-B279} have revealed that this configuration has the highest mixing rates downstream from the release nozzle.
	
	In addition to round jets, several investigations \cite{Mi2010FTaC583,Quinn1989PoFA1716,Grinstein1995PoF1483,ZAMAN1999JoFM197a} have examined jet releases through different shaped orifices of varying aspect ratios, also through flat plates. Results from these investigations have shown that asymmetric behaviours emerge, such as the \emph{axis-switching} phenomenon.  Such behaviour, and other related mechanisms, lead to increased mixing, turbulence intensity, and entrainment rates compared to round jets. In other investigations, buoyancy effects on horizontal jets \cite{L.K.2010JoFM59} have been investigated, while others \cite{Rodi1982,Carazzo2006JoFM137} have extended the survey of all available experimental data for both turbulent buoyant/pure jets and plumes to provide a quantitative study into the buoyancy effects. According to theory \cite{george1989self}, jets and plumes both have different states of partial or local self-similarity.  However, the global evolution of jets and plumes have a tendency to evolve towards complete self-similarity through a universal route, in the far-field, even with presence of buoyancy. It has also been concluded that large-scale structures of turbulence drives the evolution of the self-similarity profile, and buoyancy has an effect in exciting these coherent structures \cite{Carazzo2006JoFM137}.
	
	The influence of initial conditions on turbulent mixing and combustion performance in reactive jets, has also been of active interest in the scientific community \cite{Gutmark1989EiF248,Nathan2006496}. In last two decades, due to rapid growth in the use of hydrogen powered fuel cell vehicle, several experimental and numerical studies \cite{Chernyavsky2011IJoHE2645,Ekoto2012IJoHE17446,Houf20138179,Schefer2008IJoHE4702} have also addressed different accidental hydrogen dispersion scenarios. It is noteworthy that all aforementioned studies, as well as related previous investigations on jets or plumes, have been limited to leaks through flat surfaces, where the direction of the jet mean flow was aligned with the flow origin. All of this work has been of prime importance to determine the dispersive nature of gases, for fuel-safety purposes, for gas leaks of various hole geometries and inflow conditions.  In reality, however, accidental fuel leaks would not be limited to flows through flat surfaces.  From a practical point of view, flow patterns and dispersion of gas originating from holes in the side walls of circular pipes should also receive attention.  To date, and to the our knowledge, no such investigation has been formally published.

	In the current investigation, jets issuing from such realistic geometry were considered experimentally and numerically.  Turbulent vertical jets, flowing through a 2mm diameter round hole in the side of a 6.36 mm diameter round tube, were studied. The investigation thus considered flow through a curved surface from a source whose original velocity components were nearly perpendicular to the direction of the ensuing jets. From now on, we refer to this jet configuration as a 3D jet. This orientation permitted practical flow velocity and concentration field measurements for a realistic scenario, which were compared to axisymmetric leaks through flat surfaces accordingly.  Particle imaging velocimetry (PIV) and acetone-seeded planar laser-induced fluorescence (PLIF) were used to measure high-resolution instantaneous velocity and concentration fields, respectively, through experiment.  To compliment the experiments, large eddy simulation (LES) was also employed to model the gas dispersion.  An efficient Godunov solver was used, and coupled with adaptive mesh refinement (AMR) to provide high-resolution solutions only in areas of interest.  The fluids considered experimentally and numerically were air and helium.  Hydrogen was also considered for the numerical investigation. Thus, different fluid densities, ratio of specific heats, and buoyancy were considered accordingly.  The outer-scale flow Reynolds numbers, based on the orifice diameter, and Mach numbers of the jets ranged from 18,000 to 56,000 and 0.4 to 1.5, respectively. The purpose was to identify and characterize departures from standard axisymmetric jet conditions, and to highlight the asymmetric nature of the 3D jets, which ensued from a practical geometry arrangement. To further compare the 3D jets with axisymmetric jets, measurements were also carried out for the same physical jet conditions, and hole diameter, through a sharp-edged {orifice plate (OP)} type flat surface jet. The results obtained suggest that discrepancies exist, when applying conventional assumptions for round jets issuing through flat surfaces, to determine statistical properties associated with gas leaks from pipelines. 

\section{Methodology}

\subsection{Experimental system and techniques}
\subsubsection{Flow facility}

\begin{figure}[h!]
	\centering
	a)\includegraphics[scale=0.25]{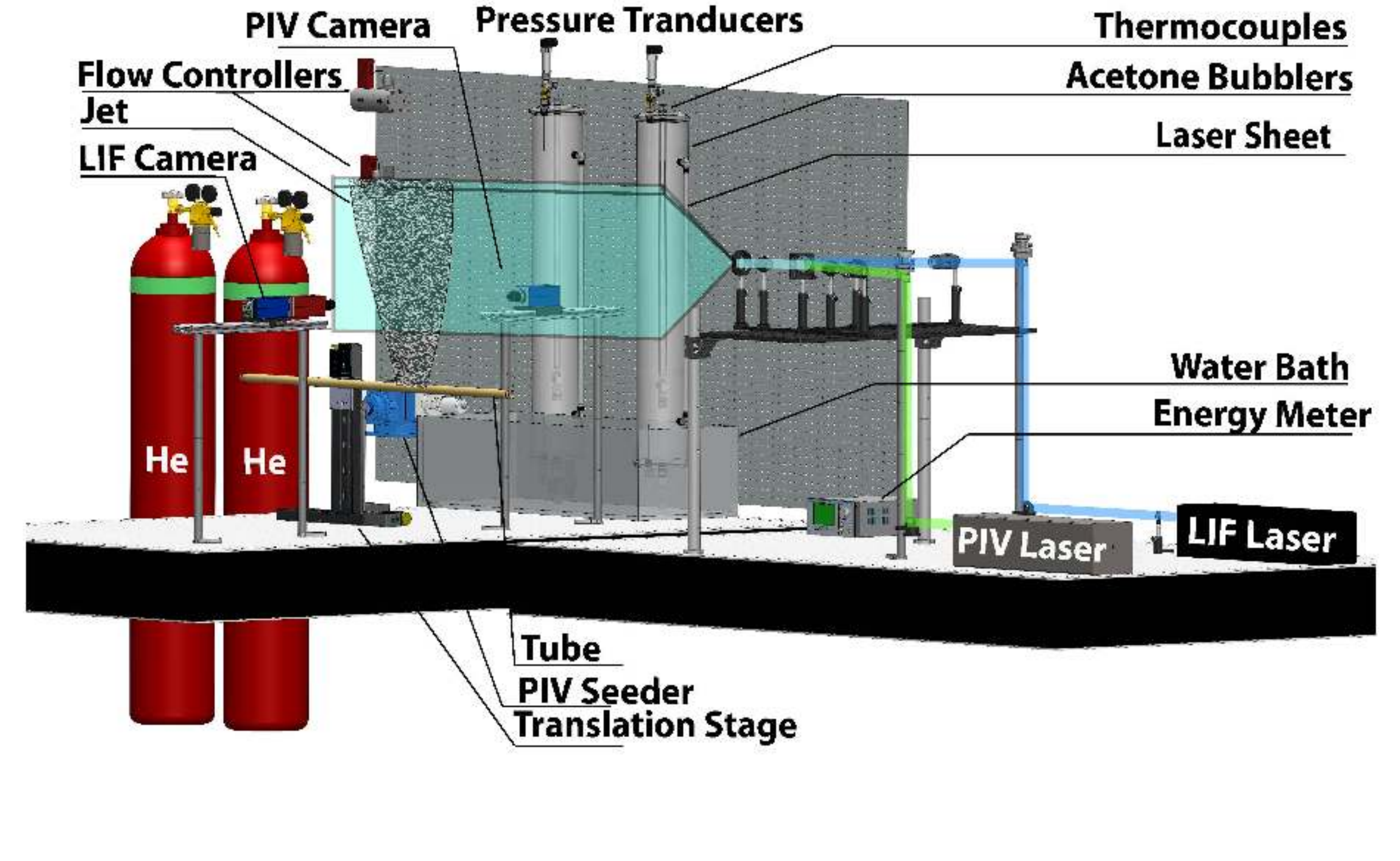}
	b)\includegraphics[scale=0.20]{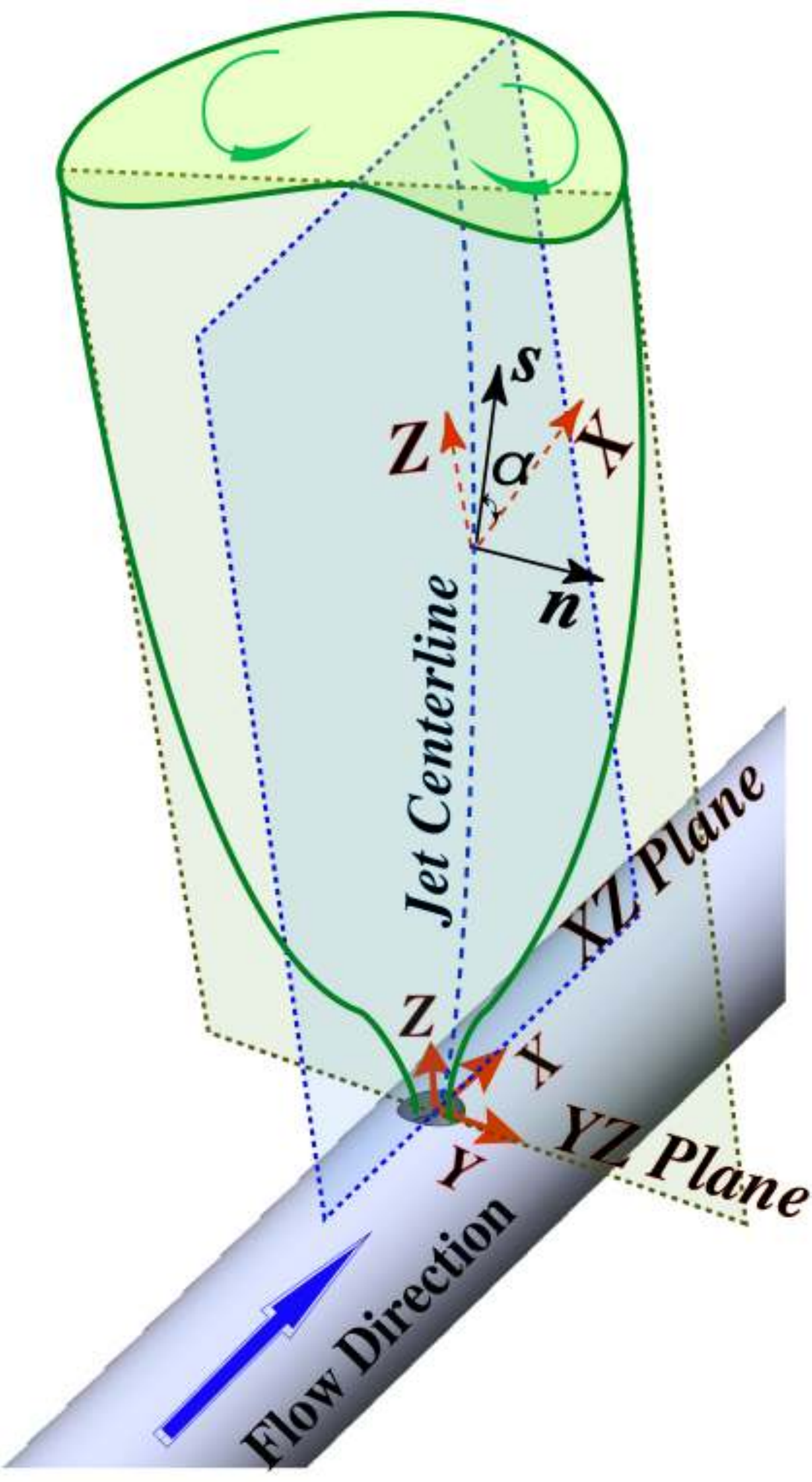}
	\caption{a) Schematic of the experimental layout. b) Illustration of the 3D jet flow experiment.}
	\label{fig.Experimental_Layout}
\end{figure}

Figure \ref{fig.Experimental_Layout}a, provides a schematic overview of the experimental setup used for this study. The experiments were conducted within a controlled stagnant environment, at room temperature and pressure ($\hat{T}_o\sim22^{\circ}$, $\hat{p}_o\sim100$ kPa).  Dry filtered air was supplied by a central flow facility, while pure scientific grade helium was supplied through compressed T-cylinders.  Flow controllers (Bronkhorst, EL-FLOW series) were used {to control} mass flow rates to the system, with a high accuracy (standard $\pm0.5\%$ of reading plus $\pm0.1\%$ full scale) and precision (within $0.2\% $ of the reading). For each experiment the test gas was passed through the PIV seeder (LaVision Aerosol Generator) at a constant pressure to ensure a consistent amount of tracer particles in all tests.  Di-Ethyl-Hexyl-Sebacate (DEHS) tracer particles were used, with a typical diameter of less than 1 \gmu m.  The test gas was also passed through two `bubbler'-type seeders.  These seeders contained liquid acetone to be used as fluorescent tracers for the PLIF.  A water bath was used to control the acetone temperature to allow acetone vapours to mix with the test gas isothermally to a saturated state.  In all experiments, the test gas, which contained both tracers, was consistently $\sim10\%$ saturated gas-acetone mixture by volume.  All mixing procedures were also controlled by mass flow controllers.  The mixing was monitored by pressure transducers and thermocouples at several locations within the system. Isothermal and isobaric conditions were thus ensured {in all experiments.} After the test gas was mixed and seeded with the PIV and PLIF tracers, the flow entered the test section of the tube.\par

Figure \ref{fig.Experimental_Layout}b {illustrates} the jet flow evolution from the tube orifice. To capture the three-dimensionality of the jet, measurements {were} obtained on two different planes (denoted $x$-$z$ and $y$-$z$) for each experiment, as indicated.  Also {shown} in the figure is the jet centreline, which acts as a reference from which measurements are later obtained in the $x$-$z$ plane.  Owing to potential deviation of the jet from the orifice axis ($z$-axis), the jet centreline tangent and normal lines are shown as $s$ and $n$ coordinates in the figure, respectively. 

In order to compare the evolution of the different test gases, the averaged momentum flux $(\overline{\hat{\rho} \hat{u}})_{\textrm{flux}}$ at the jet exit was measured and matched for all test cases, as suggested in previous studies \cite{Chen1980,Panchapakesan1993JoFM225}. This matching was achieved, {iteratively}, by varying the volumetric flow rate ($\hat{Q}$) in the system.  Here,  $(\overline{\hat{\rho} \hat{u}})_{\textrm{flux}}$ was calculated by first obtaining the time-averaged jet exit velocity, which was measured at the closest vertical distance, $z\simeq0$, to the orifice.  The momentum flux was then calculated from
\begin{equation}
(\overline{\hat{\rho} \hat{u}})_{\textrm{flux}}=2\pi\int_{0}^{\hat{D}/2} {\hat{\rho}_{{j}}  \overline{\hat{u}}^{2}(\hat{r})  \diff{\hat{r}}}
\label{eqn.momentum_flux_experiment}
\end{equation}
where the subscript `$j$' refers to the conditions at the nozzle, the over-line refers to a time-averaged quantity, and the hat refers to a dimensional quantity.  Here $\hat{D}=2$ mm was the diameter of the orifice.  $\rho$ and $r$ refer to density and radius, respectively. Table \ref{tab.Flow properties} shows the flow properties used in this study, for both the 3D and OP jet configurations. The flows were characterized by the outer-scale Reynolds number, $Re_{\delta}= \overline{\hat{u}}_{j}\delta/\hat{\nu}_{\infty}$.  Here, $\hat{\nu}_{\infty}$ is the ambient fluid kinematic viscosity and   $\delta$ is the width of the mean axial velocity profile, evaluated from limits of 5\% of the centreline velocity at $z\simeq0$.

\begin{table}[h!]
	\caption{Flow properties}
	{\small
		\begin{tabular}{|l|l|l|l|l|l|l|l|l|}
			\hline
			Jet &  $\hat{Q}$ & $\overline{\hat{u}}_{{j}}$ &$\hat{\rho}_{{j}}$ &$\hat{\nu}_{{j}}$  &$ (\overline{\hat{\rho} \hat{u}})_{\textrm{flux}} $  & $Ma$ &  $Fr$ & $Re_{\delta}$\\
			  &   [L/min] & [m/s]  & [Kg/$\textrm{m}^{3}$] &  [$\textrm{m}^{2}$/s]  & [N]  &   & & \\						
			\hline
			3D Air&15&147.5 & 1.17 & $1.54 \times 10^{-5} $&0.1018 & 0.43 & - &19,000 \\
			\hline
			OP Air&15&127.6 & 1.17 & $1.54 \times 10^{-5} $&0.0762 & 0.37 & - &16,500 \\
			\hline
			3D He&35& 399.7& 0.165 &$1.21 \times 10^{-4} $  &0.1048& 1.2 & 1144 &51,500 \\			
			\hline
			OP He&35& 341.9& 0.165 &$1.21 \times 10^{-4} $  &0.0767& 1.0 & 978 &44,200 \\			
			\hline
	\end{tabular}}
	\label{tab.Flow properties}
\end{table}

\subsubsection{Velocity measurements}
For the PIV technique, a two-dimensional cross-section of the seeded flow was illuminated using a pulsed laser.  The scattered light from the tracer particles was recorded using a charge-coupled device (CCD) camera to obtain {the instantaneous locations of the particles. The velocity field was then obtained for each pair of instantaneous particle location images accordingly.}  A dual-head Nd: YAG pulsed laser (New Wave's SOLO III  15 HZ) was used to provide a {light source at a wavelength of 532 nm.  The optical system was} designed to create a light sheet with an approximate height {of} 5 cm and thickness {of} 1 mm. The PIV CCD camera was equipped with a Nikon Micro-NIKKOR 60 mm lens, and the lens aperture was kept at {(f4)}.  To suppress background light, a 532 nm bandpass filter with the full width at half maximum (FWHM) of 10 nm was attached {to} the camera lens. The field of view of the camera was a 40$\times$30 mm {rectangle} with an approximate pixel size of 6.5 \gmu m {in physical space}. Following the procedure of Su \cite{Su2003JoFM1}, we estimate this resolution to be comparable to the finest scales of the flow, with respect to the Nyquist criterion.  The {instantaneous particle location} images were obtained at {a} frequency of 10 Hz and each pair of images was processed using LaVision DaVis 8.4 software to calculate the instantaneous velocity fields. {The corresponding velocity fields were thus compiled at a temporal frequency of 5 Hz}.  This process was followed by {applying} a multi-pass spatial resolution improvement algorithm.  With each pass, the interrogation window size, {corresponding to a single calculated velocity vector field}, was decreased from 32$\times$32 to 16$\times $16 pixels, with a 75\% overlap {between the windows} in the horizontal and vertical directions. For each {experiment case}, a total of $N=500$ {velocity field} images were acquired for statistical averaging.

\subsubsection{Concentration measurements}

{To measure the concentration of the jet gas in the flow field, we applied PLIF.  This is a non-intrusive, spatially resolved, laser diagnostic technique. Similar to PIV, the application of PLIF, in our case,} relied on a pulsed laser sheet to illuminate a two-dimensional section of {the} flow field. The wave length of the laser {light} was tuned to excite the {tracer} acetone molecules, which were artificially {introduced to} the flow field. {An} ultraviolet wavelength {was} used to produce electronic excitation.  A fraction of {the} excited {tracer} molecules {emitted photons} while simultaneously returning to the equilibrium state. This {resulted in a} measurable fluorescence {signal} from the tracers, which was captured here by {a} CCD camera.  In this study, acetone was chosen as the tracer over other alternatives for {several} reasons. First, a high vapour pressure at room temperature absorbs over a wide band of wavelengths (225-320 nm) and emits fluorescence on an even wider broadband of wavelengths (350-550 nm).  Acetone also has a short fluorescence lifetime ($\sim2$ ns), {negligible oxygen quenching on fluorescence signal, and its} fluorescence signal in isothermal and isobaric flows is known to be linear with laser power and concentration\cite{Lozano1992EiF369}.  A Pulsed Nd: YAG laser (Spectra-Physics INDI-40-10-HG) {was} used to provide a stable 266 nm wavelength ultraviolet light in order to excite the acetone molecules. In order to {obtain a} linear fluorescence regime,  optical lenses were used to create a light sheet with an approximate height of 5 cm {and} a thickness of 350 \gmu m. This ensured that saturation of the fluorescence signal did not occur until laser energy per pulse reached $\sim10.4$ J, which was well above the maximum laser energy output (55 mJ) used in this study.

\begin{figure}[h!]
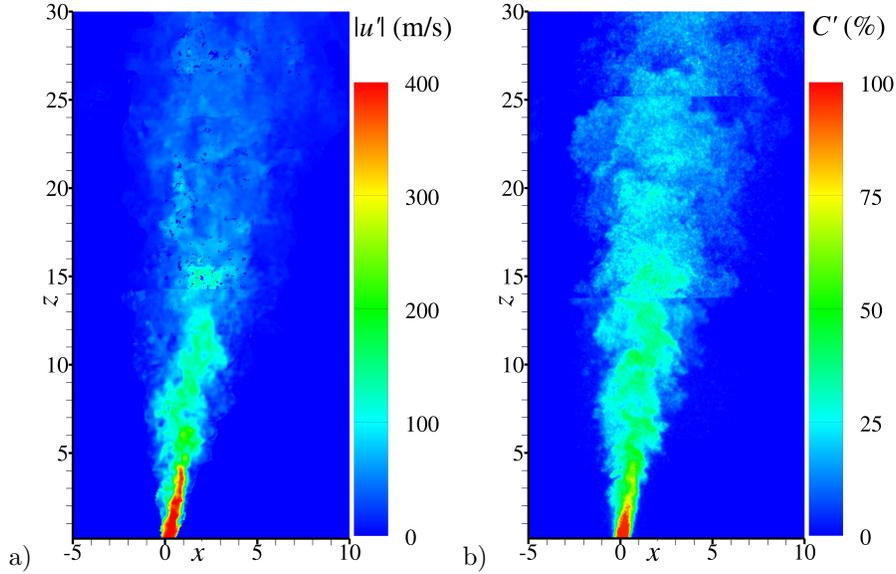

	\centering
	a)\includegraphics[scale=0.04,trim={0.5cm 0.5cm 0.5cm 0.5cm},clip]{Figure2a.pdf}
	b)\includegraphics[scale=0.04,trim={0.5cm 0.5cm 0.5cm 0.5cm},clip]{Figure2b.pdf}\\
	\caption{Instantaneous a)  velocity and b) concentration fields obtained from Helium \newline
		\hspace{\linewidth}  3D jet in $x$-$z$ plane.}
	\label{fig.Instantaneous_plots}		
	\end{figure}

The PLIF CCD camera was equipped with an intensifier unit, which was sensitive to the ultraviolet spectrum, in order to increase gain and gating capability. The camera was also equipped with a Nikon Micro-NIKKOR 105 mm lens. {The} aperture was kept open at {(f2.8)}, and a 378-nm UV bandpass filter with FWHM of 140 nm was used.  The camera field of view for all cases corresponded to a 38$\times$28 mm window with an approximate pixel size of 6.5 \gmu m. Before each experiment, sets of background and laser sheet images were {taken to} determine the averaged background and cross-sectional laser beam intensity distributions.  These {time-}averaged images were later subtracted from each raw {PLIF} image in order to {isolate} the fluorescence signal. In order to account for the laser beam energy fluctuations, all images {were} normalized by the {amount of the} laser energy per pulse, which was measured using a laser energy meter.  The images were taken at {a} frequency of 5 Hz and then processed using LaVision DaVis 8.4 software. For each {experimental case}, a total of $N=500$ images were acquired to determine the time-averaged concentration ($\overline{C}$) fields.  Figure \ref{fig.Instantaneous_plots} shows an example of the instantaneous velocity and concentration fields, for the helium 3D jet in the $x$-$z$ plane, which were obtained simultaneously.

\subsection{Numerical techniques}
\subsubsection{Governing equations}

For flows which are turbulent and compressible, the gas dynamic evolution is governed by the compressible Navier-Stokes equations.  In order to account for the full spectrum of {turbulent} scales, resulting from large flow velocities with high Reynolds numbers ($Re$), the unresolved scales of the governing equations are filtered and modelled through the {large eddy simulation (LES)} approach \cite{Moin2002}. The LES-filtered conservation equations for mass, momentum, and energy (sensible + kinetic) of a calorically perfect fluid system are given below in Eqs.\ \eqref{eqn.LESmass2}-\eqref{eqn.LESenergy2}, respectively.  Also, a transport equation \eqref{eqn.LESspecies} is included to describe the evolution of mass fraction ($Y$) associated with the jet gas.  The governing equations are supplemented by a one-equation localized kinetic energy model \cite{Chakravarthy2001}, given by Eq.\ \eqref{eqn.LESke}, {which} describes the transport, production, and dissipation of the subgrid kinetic energy (${k}^{sgs}$) associated with subgrid velocity fluctuations.  Finally, the equations of state are given by \eqref{eqn.EOS}.  The equations presented here are given in non-dimensional form, where the various properties are normalized by the reference quiescent state. Favre-average filtering is achieved by letting $\tilde{f} = \overline{\rho f} / \bar{\rho}$, where $f$ represents one of the {state variables and} $\rho$, $p$, $e$, $T$, and $\boldsymbol{u}$ refer to the density, pressure, specific sensible + kinetic energy, temperature, and velocity vector, respectively.  Other {relevant} properties are the ratio of specific heats, $\gamma$, the kinematic viscosity, $\nu$, the resolved shear stress tensor, $\bar{\bar{\boldsymbol{\tau}}}$, the Prandtl number, $Pr$, and the Schmidt number, $Sc$.  Also, subgrid contributions due to buoyancy have been accounted for, where $\boldsymbol{g}$ is the gravitational acceleration.

\small\begin{equation}
{\frac{\partial \bar{\rho}}{\partial {t}}} + {\boldsymbol{\nabla} \cdot (\bar{\rho} \tilde{{\boldsymbol{u}}})} = 0
\label{eqn.LESmass2}
\end{equation}%
\begin{equation}
{\frac{\partial (\bar{\rho} \tilde{\boldsymbol{u}})}{\partial {t}}} + {\boldsymbol{\nabla} \cdot (\bar{\rho} \tilde{\boldsymbol{u}} \tilde{{\boldsymbol{u}}})} + {\nabla \bar{p}} - {\boldsymbol{\nabla} \cdot {\bar{\rho}(\nu + \nu_{t})} \biggl( \nabla \tilde{\boldsymbol{u}} + (\nabla \tilde{\boldsymbol{u}})^T - \frac{2}{3}( \boldsymbol{\nabla} \cdot \tilde{{\boldsymbol{u}}} ) \boldsymbol{I} \biggr)} = \bar{\rho}\boldsymbol{g}
\label{eqn.LESmomentum2}
\end{equation}%
\begin{equation}
{\frac{\partial (\bar{\rho} \tilde{e})}{\partial {t}}} + {\boldsymbol{\nabla} \cdot \biggl((\bar{\rho}\tilde{e}+\bar{p})\tilde{{\boldsymbol{u}}} - \tilde{{\boldsymbol{u}}} \cdot \bar{\bar{\boldsymbol{\tau}}}\biggr)} - { \boldsymbol{\nabla} \cdot \biggl( \bar{\rho}\biggl(\frac{\gamma}{\gamma - 1} \biggr) \biggl(\frac{\nu}{Pr} + \frac{\nu_t}{Pr_t}\biggr) \nabla \tilde{T} \biggr)} = \bar{\rho}\tilde{\boldsymbol{u}}\cdot\boldsymbol{g}
\label{eqn.LESenergy2}
\end{equation}
\begin{equation}
\frac{(\partial \bar{\rho} \tilde{Y})}{\partial t} + \boldsymbol{\nabla} \cdot (\bar{\rho}\tilde{\boldsymbol{u}}\tilde{Y}) - \boldsymbol{\nabla} \cdot {\biggl(\bar{\rho} \biggl(\frac{\nu}{Sc} + \frac{\nu_t}{Sc_t}\biggr) \nabla \tilde{Y}\biggr)} = 0
\label{eqn.LESspecies}
\end{equation}
\begin{equation}
{\frac{\partial (\bar{\rho} {{k}^{sgs}})}{\partial {t}}} + {\boldsymbol{\nabla} \cdot (\bar{\rho} \tilde{{\boldsymbol{u}}} {{k}^{sgs}})} - {\boldsymbol{\nabla} \cdot \biggl({\frac{\bar{\rho} \nu_{t}}{Pr_t} \nabla {{k}^{sgs}}} \biggr)} = -\frac{\nu_t}{Pr_t}\nabla\bar{\rho}\cdot\boldsymbol{g} + {\dot{P}} - {\bar{\rho} \epsilon}
\label{eqn.LESke}
\end{equation}%
\begin{equation}
\tilde{e} = {\frac{{\bar{p}}/{\bar{\rho}}}{(\gamma - 1)}} + {\frac{1}{2} \tilde{\boldsymbol{u}} \cdot \tilde{\boldsymbol{u}}} + { {{k}^{sgs}}} \;\;\;\;\;\;\;\; \mbox{and} \;\;\;\;\;\;\;\;  \frac{\bar{p}}{\bar{\rho}} = {R\tilde{T}}
\label{eqn.EOS}
\end{equation}%
\normalsize
The various state variables have been normalized such that
\small\begin{equation}
\rho = \frac{\hat{\rho}}{\hat{\rho_o}}, \; \boldsymbol{u} = \frac{\hat{\boldsymbol{u}}}{\hat{c_o}},  \; p = \frac{\hat{p}}{\hat{\rho_o} {\hat{c_o}}^2} = \frac{\hat{p}}{\gamma \hat{p_o}}, \; T = \frac{\hat{T}}{\gamma \hat{T_o}}, \; x = \frac{\hat{x}}{\hat{D}}, \; t = \frac{\hat{t}}{\hat{D} / \hat{c_o}},  \; R=\frac{\hat{R}}{\hat{R}_o}=\frac{1/\hat{\overline{W}}}{1/\hat{W}_o},
\label{eqn.nonDim1}
\end{equation}\normalsize
where the subscript `o' refers to the reference state, the hat superscript refers to a dimensional quantity, $\boldsymbol{I}$ is the identity matrix, $c$ is the speed of sound, $W$ is the molecular weight, and $D$ is the reference diameter of the orifice through which the gas exits the pipe.  The {rates} of production and dissipation of $k^{sgs}$ are given by
\begin{equation}
\dot{P} = \bar{\rho}\nu_{t} \biggl( \nabla \tilde{\boldsymbol{u}} + (\nabla \tilde{\boldsymbol{u}})^T - \frac{2}{3}( \boldsymbol{\nabla} \cdot \tilde{{\boldsymbol{u}}} ) \boldsymbol{I} \biggr) \cdot (\nabla \tilde{\boldsymbol{u}})\ \ \ \ \ \textrm{and}\ \ \ \ \ \epsilon = {\pi\biggl(\frac{2{k}^{sgs}}{3C_\kappa}\biggr)^{3/2}}/{\bar{\Delta}}.
\label{eqn.turbulentViscosity}
\end{equation}%
Next, a Smagorinsky-type model is applied to describe $\nu_t$ in terms of $k^{sgs}$ through 
\begin{equation}
\nu_{t} = \frac{1}{\pi}\biggl(\frac{2}{3C_\kappa}\biggr)^{3/2}\sqrt{k^{sgs}}\bar{\Delta}.
\label{eqn.turbulentViscosity}
\end{equation}%
Here, $C_{\kappa}$ is the \emph{Kolmogorov constant}, whose value is set to a typical value of $C_\kappa=1.5$.  For simplicity, the LES filter size, $\bar{\Delta}$, is assumed to be equal to the (local) minimum grid spacing.  It is noted, however, that this assumption may introduce some errors at fine-coarse cell interfaces when {using} {adaptive mesh refinement (AMR)} \cite{Vanella2008}.  Finally, for the helium case, owing to differences in $\gamma$, Eq.\ \eqref{eqn.LESspecies} is replaced with
\begin{equation}
\frac{(\partial \bar{\rho} \tilde{G})}{\partial t} + \boldsymbol{\nabla} \cdot (\bar{\rho}\tilde{\boldsymbol{u}}\tilde{G}) - \boldsymbol{\nabla} \cdot {\biggl(\bar{\rho} \biggl(\frac{\nu}{Sc} + \frac{\nu_t}{Sc_t}\biggr) \nabla \tilde{G}\biggr)} = 0
\label{eqn.LESgamma}
\end{equation}%
where
\begin{equation}
\tilde{Y}=\frac{\tilde{G}-{G}_{\textrm{air}}}{{G}_{\textrm{He}}-{G}_{\textrm{air}}}\ \ \ \ \ \textrm{and}\ \ \ \ \ {G}=\biggl(\frac{1}{{\gamma}-1}\biggr).
\end{equation}%
Although {the} conservative form {used in this method} is known to introduce pressure oscillations, which originate from material interfaces \cite{Shyue1998}, it is necessary to ensure the correct mathematical representation of the diffusion process.  While non-conservative approaches {do not exhibit such pressure oscillations} \cite{Shyue1998}, they can also converge to {physically} incorrect solutions with respect to diffusion \cite{Hou1994}.  For practical purposes, $\gamma$ is evaluated from $\tilde{G}$ directly, as no suitable alternative exists in the LES framework.

\subsubsection{Domain and model parameters}
\begin{figure}[ht]
	\centering
	\vspace*{10pt}
	\includegraphics[scale=0.667]{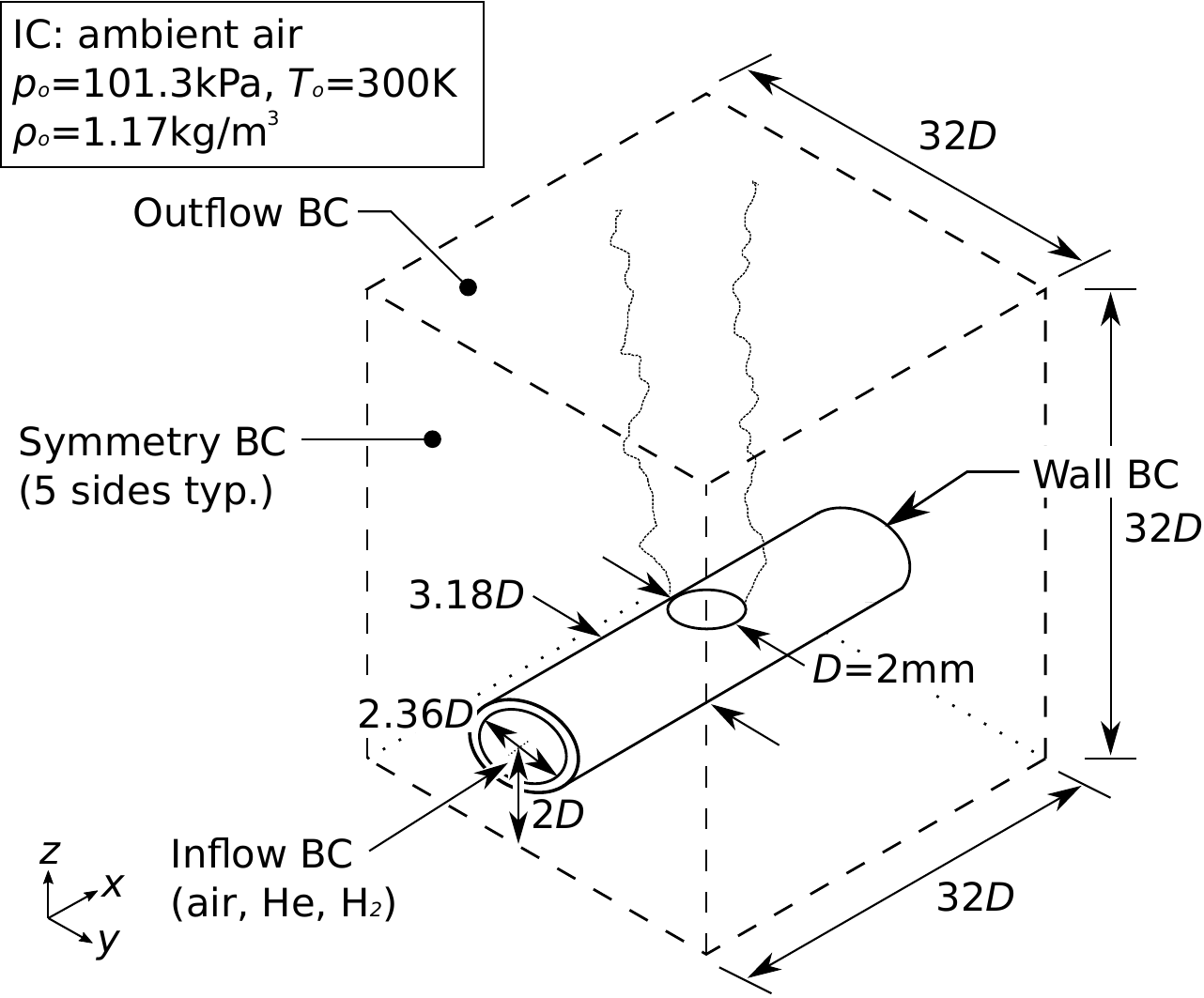}
	\caption{Computational domain with initial and boundary conditions (not to scale).}
	\label{fig.domain}
\end{figure}

The numerical domain containing the pipe and jet configuration is shown in Fig.\ \ref{fig.domain}.  The pipe had an outer diameter of $3.18D$ (6.36mm) with a wall thickness of $0.41D$ (0.82mm).  The hole, through which gas escaped, had a diameter of $\hat{D}=2$mm.  The domain itself spanned $32D$ in each direction.  The inlet boundary condition (BC) was imposed on one side of the pipe, which used a digital filtering generation method \cite{Klein2003} to generate the appropriate second order turbulence characteristics according to well-documented experimental measurements of turbulence in pipe flow \cite{Eggels1994}.  A wall BC was imposed on the other side of the pipe, which causes the flow to stagnate within the pipe.  This was consistent with the experiments.  The top BC of the domain was a pressure outlet type.  The remaining 5 BCs were symmetry type slip walls, and were sufficiently far away from the jet to prevent interference.  The {simulations were initialized with} air at ambient conditions ($\hat{T}_o=300$K and $\hat{p}_o=101.3$kPa) {throughout the domain}.

To be consistent with the experiments, the average momentum flux $(\overline{\rho u})_{\textrm{flux}}$ was matched for all simulations.  To achieve this, the inlet pressure was varied, through trial and error, to obtain the desired $(\overline{\rho u})_{\textrm{flux}}$ and average flow velocity ($\overline{u}_j$).  Since three-dimensional information was available, the instantaneous $({\rho u})_{\textrm{flux}}$ was {monitored} directly on the $x$-$y$ plane corresponding to the hole location on the outer edge of the pipe, at $z=0$.  In this case,
\begin{equation}
(\rho u)_{\textrm{flux}}=\int_{z=0}{\rho u_z u_z \diff A}.
\end{equation}%
The resulting time-averaged $(\overline{\rho u})_{\textrm{flux}}$, jet velocity ($\overline{u}_j$), and other relevant fluid properties are given in Table \ref{tab:tab}.  The transport properties have been evaluated at equimolar conditions of ambient air and the jet gas, and were assumed constant for simplicity.  For the turbulent transport properties of all {three} jets, $Sc_t=0.7$ and $Pr_t=0.8$ were also assumed constant.  Finally, for each jet, a total of $N=500$ (for air) or $N=1500$ (for $H_e$/$H_{2}$) time steps were processed for statistical averaging, once a quasi-steady jet was established.  This corresponded to sampling over eddy turnover times of {$\tau=450$, 1800, and 2040} for air, helium, and hydrogen, respectively.

\begin{table}[ht!]
	\caption{Model Parameters.}
	{\small
		\begin{tabular}{|l|l|l|l|l|l|l|l|l|l|}
			\hline
			 Jet & $\overline{\hat{\rho}}_j$  & $\overline{\hat{u}}_j$  & $Ma$ & {$Re$} &  $(\overline{\hat{\rho} \hat{u}})_{\textrm{flux}} $ & $\gamma$  & {$\hat{\nu}$  } & $Pr$  & {$Sc$}  \\

			  & [kg/$\textrm{m}^3$] & [m/s] & & & [N] &  &  [$\textrm{m}^2/\textrm{s}$]  &   &   \\

			\hline
			air  & 1.17 & 141.7 & 0.4 & 17,824 & 0.0335 & 1.4 & $1.59 \times 10^{-5}$ & 0.714 & 0.707  \\
			\hline
			He  & 0.164 & 368.1 & 1.1 & 38,545 & 0.0317 & 1.67 & $1.91 \times 10^{-5}$ & 0.607 & 0.626   \\
			\hline
			$\textrm{H}_2$  & 0.082 & 528.4 & 1.5 & 55,915 & 0.0328 & 1.4 & $1.89 \times 10^{-5}$ & 0.556 & 0.609   \\
			\hline
	\end{tabular}}
	\label{tab:tab}
\end{table}

\subsubsection{Numerical implementation}
{In order to solve the system of equations} \eqref{eqn.LESmass2} through \eqref{eqn.LESke}, {an efficient second order accurate \emph{exact} Godunov compressible flow solver} \cite{Gottlieb1988}, {which features a symmetric monotonized central flux limiter} \cite{vanLeer1977}, {was applied to treat the advection terms consisting of potentially different $\gamma$ values on each cell interface. The diffusive terms were handled explicitly in time and spatially discretized using second order accurate central differences}.  {Structured Cartesian grids were applied in order to take advantage of AMR} \cite{Falle1993} {for increased efficiency.  The grid was refined, on a per cell-basis, in regions close to the physical pipe, and also where the jet gas mass ($\bar{\rho} \tilde{Y}$) changed by more than 0.01\% locally between existing grid levels.  The jet was refined to a minimum grid size of $D/16$ up to 10$D$ downstream from the orifice in order to capture fine scale turbulent motions in the near field.  Beyond 10$D$ downstream, the jet was only refined to a minimum grid size of $D/8$.  Finally, once a cell was refined, it remained refined for the duration of the simulation.  This {approach} avoided complications which could arise due to cell-derefinement and re-refinement \cite{Mitran2009}.  Figure \ref{fig:grid_topology} shows a typical grid topology that develops as a jet evolves in time.  The figure also indicates the locations of each grid level (G) in a portion the flow field.

\begin{figure}[h!]
	\centering
	\includegraphics[scale=0.5]{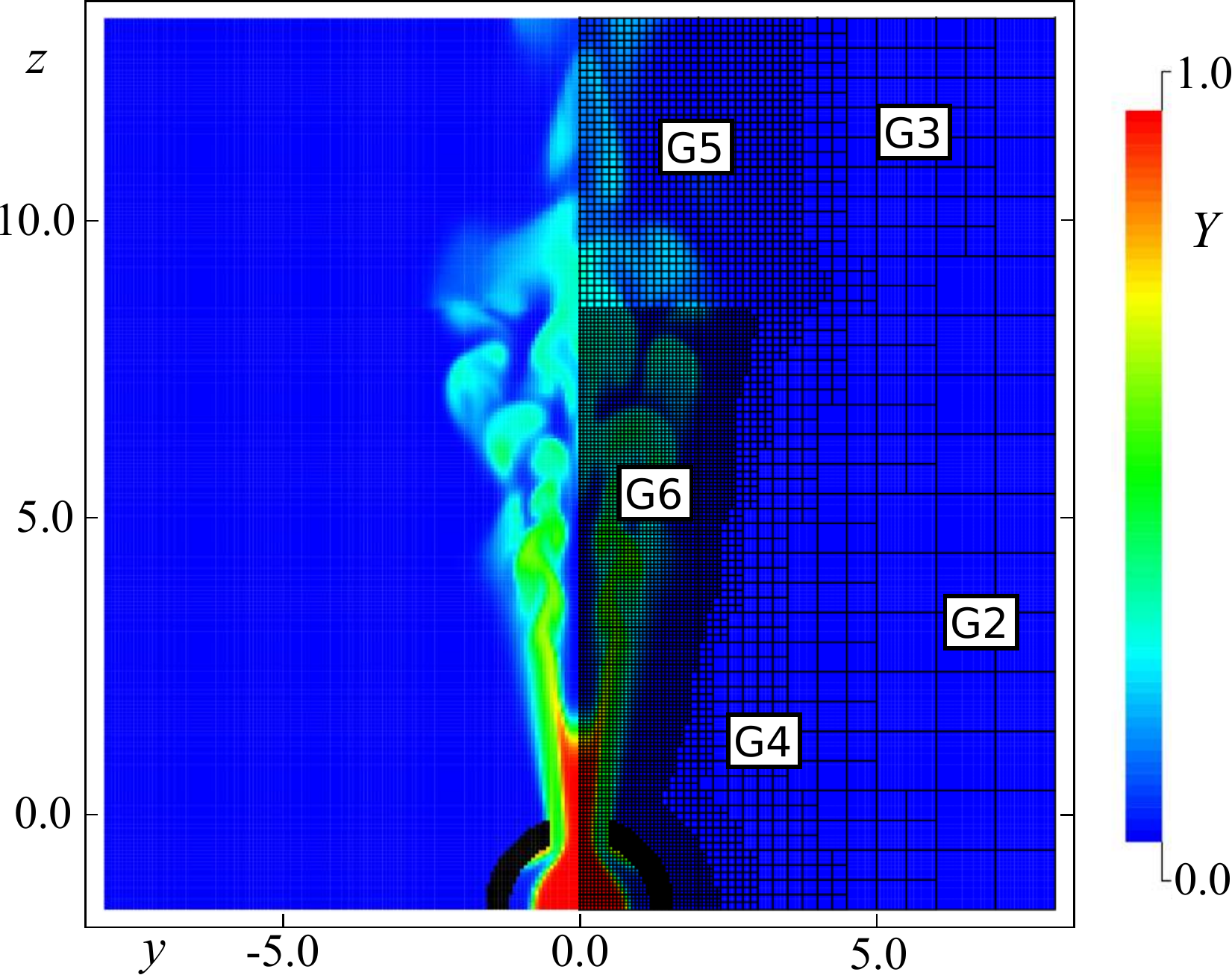}
	\caption{Instantaneous air jet showing the mass fraction $Y$ on the left and resulting grid topology, showing the locations of various grid levels (G2-G6), on the right.  Note: The base grid G1 is always refined to at least grid level G2, everywhere.}
	\label{fig:grid_topology}
\end{figure}

\subsubsection{Resolution study}
A resolution study was performed for air jet simulations at three different {spatial} resolutions. The minimum grid sizes were varied from $\bar{\Delta}=D/4$ to $\bar{\Delta}=D/16$.  Corresponding instantaneous evolutions of the jet gas mass fraction ($Y$) are presented in Fig. \ref{fig.Res_centreline}a for each resolution.  It was found that the lowest resolution, $\bar{\Delta}=D/4$, was not able to resolve any turbulent motions downstream from the orifice, and hence there was minimal jet spreading observed.  As the resolution was increased to $\bar{\Delta}=D/8$, turbulent motions were captured in the far field, {which} lead to significantly more jet spreading, and unsteadiness, as expected.  When the minimum grid size was set to $\bar{\Delta}=D/16$, small-scale turbulent motions were captured closer to the orifice, which caused a shortening of the potential jet core region owing to increased turbulent mixing.

Figures \ref{fig.Res_centreline}b and c show the jet trajectories and velocity decay rates for all three cases.  In order to measure the trajectory of each jet, whose deflection from the vertical ($z$) axis was observed previously in Fig.\ref{fig.Instantaneous_plots} , the $(x,y)$ locations of the maximum velocity magnitude ($\lvert \boldsymbol{\overline{u}} \rvert_{\textrm{max}}(z)$) were determined at discreet heights along the $z$-axis.  Also shown are the computed centre of mass locations (C.M.) for each simulation.  The C.M., as a function of height ($z$), was determined by extracting $x$-$y$ slices at each discreet heights along the $z$-axis, and evaluating the centroid associated with the average mass flux of the jet through each slice.  For a given $z$ location,
\begin{equation}
x_{\textrm{C.M.}} = \frac{\int{\overline{(\rho u_zY)}x\diff x\diff y}}{\int{\overline{(\rho u_zY)}\diff x\diff y}}\ \ \ \ \ \textrm{and}\ \ \ \ \ y_{\textrm{C.M.}} = {\frac{\int{ \overline{(\rho u_zY)}y\diff x\diff y}}{\int{\overline{(\rho u_zY)}\diff x\diff y}}}.
\end{equation}
The velocity decay rates, along the jet centre-lines, have been determined from $\lvert \boldsymbol{\overline{u}} \rvert_{\textrm{max}}(z)$.  {In general, increasing the resolution lead to increased deflection about the $z$-axis, in both $\lvert \boldsymbol{\overline{u}} \rvert_{\textrm{max}}(z)$ and the C.M., as observed in} Fig,\ \ref{fig.Res_centreline}b. {It should also be noted that the discreet jumps in the $\lvert \boldsymbol{\overline{u}} \rvert_{\textrm{max}}(z)$ locations were equal to the grid spacing associated with the corresponding resolution.  Such discreetness was not observed for the} C.M.\ {locations since the averaging process was performed across entire $x-y$ planes.  This resulted in smoother trajectories for the computed} C.M.\ {locations.  In} Fig.\ \ref{fig.Res_centreline}c, {increasing the resolution lead to an earlier decay in centreline velocity.}  This behaviour in the jet velocity decay resulted from higher resolution of fine scale turbulent motions near the orifice, which influenced earlier breakup of the potential core, thus slowing the motion of the jet.  It should be noted, however, that the actual rate of velocity decay, in the far field, was found to be the same for both the $\bar{\Delta}=D/8$ and $\bar{\Delta}=D/16$ resolutions.  These far field velocity decay rates were determined by the slopes of lines of best fit beyond $z>15D$, obtained from linear regression, as shown in Fig.\ \ref{fig.Res_centreline}c.

\begin{figure}
	\centering
	a)\includegraphics[scale=0.28]{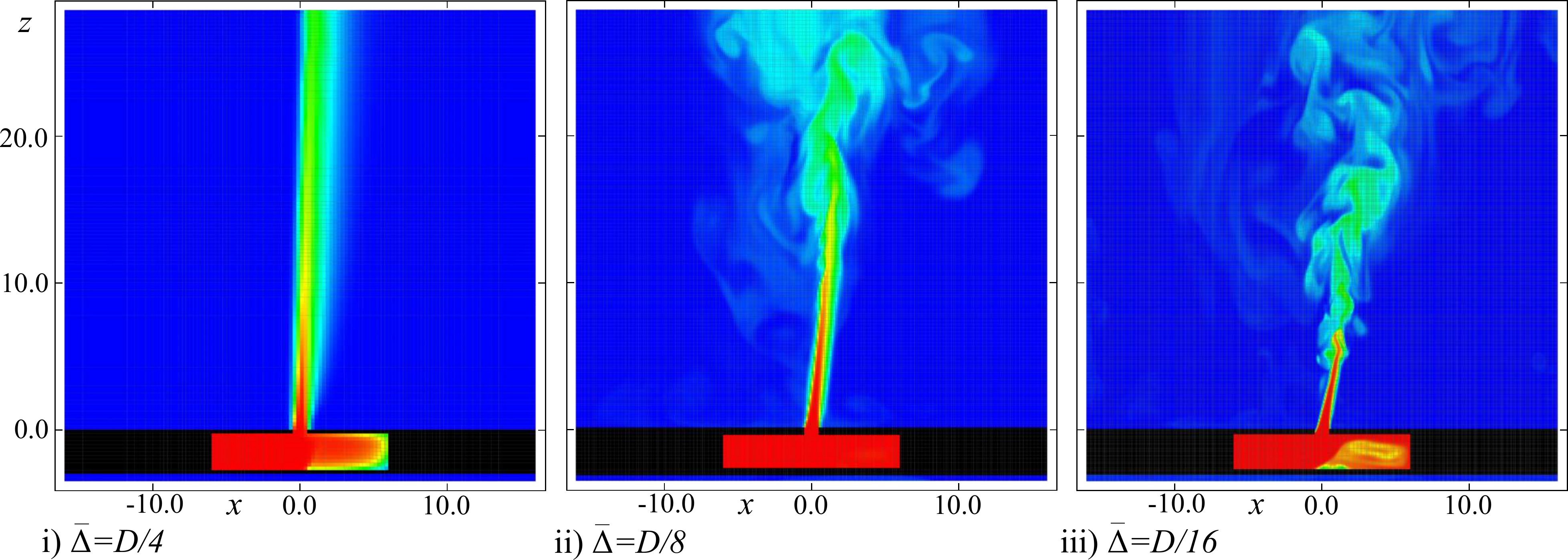}
	b)\includegraphics[scale=0.245]{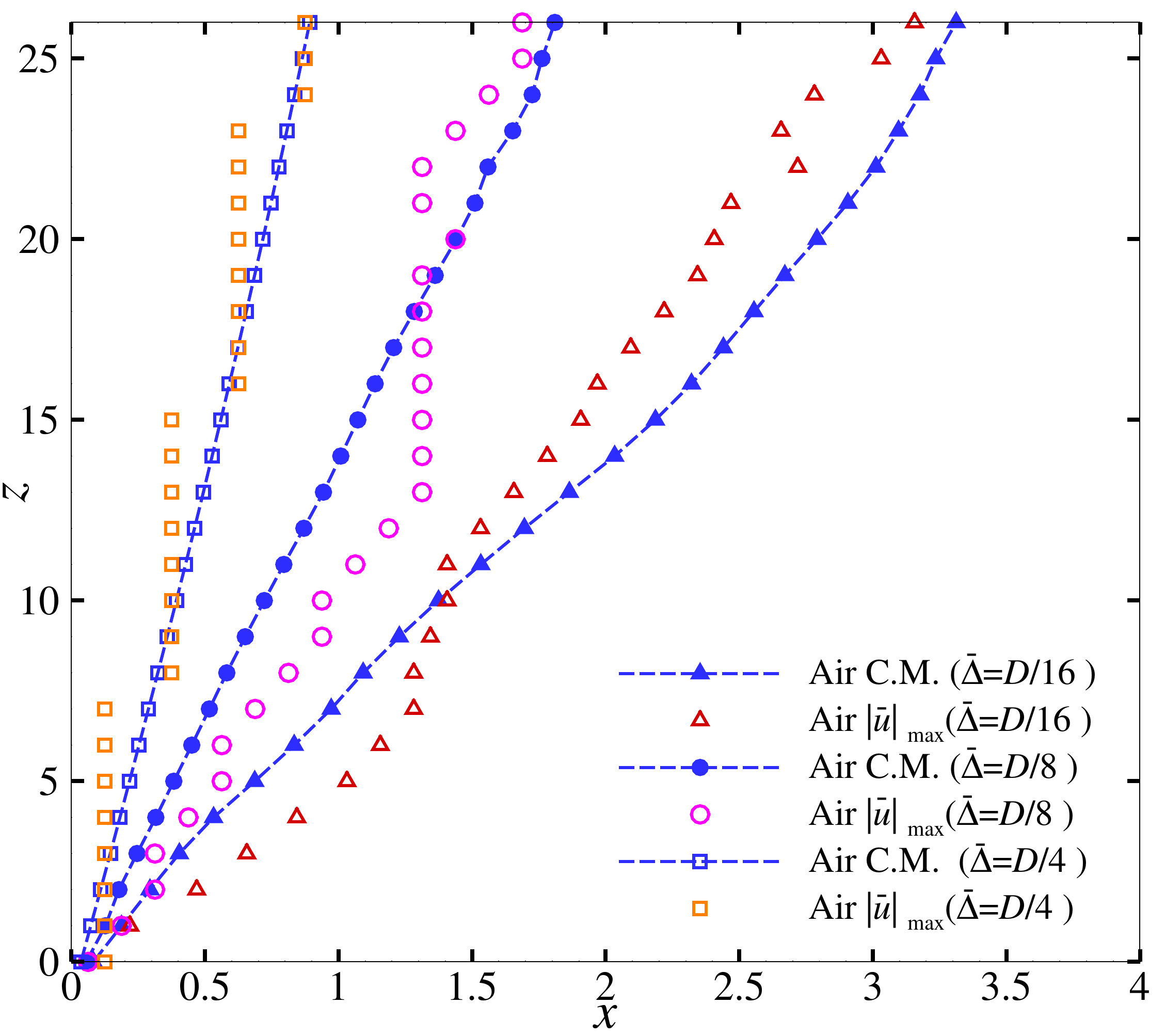}
	c)\includegraphics[scale=0.245]{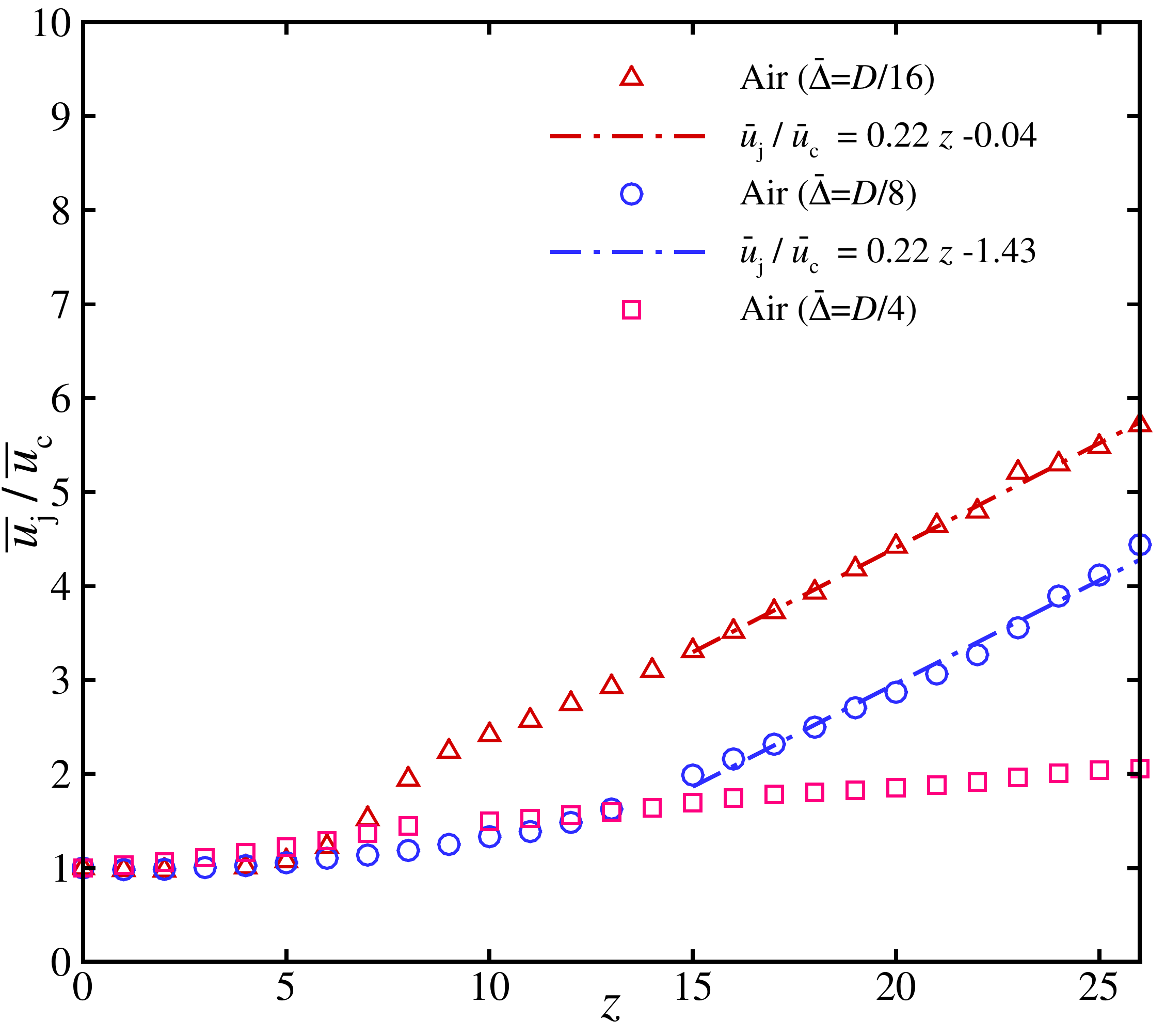}
	
	\caption{a) Instantaneous mass fraction field ($Y$) for air at different resolutions.  b) Jet centre-lines taken along the location of maximum velocity ($\lvert \boldsymbol{\overline{u}} \rvert_{\textrm{max}}(z)$), and also the centre of mass (C.M.) locations, obtained for air at different resolutions. c) Jet velocity decay rates for air simulations at different resolutions.}
	\label{fig.Res_centreline}
\end{figure}

\section{Results}

\subsection{Time-averaged flow fields}

\begin{figure}[h!]
	\centering
	\raggedright \underline{\textbf{air:}}\\
	\includegraphics[scale=1.0]{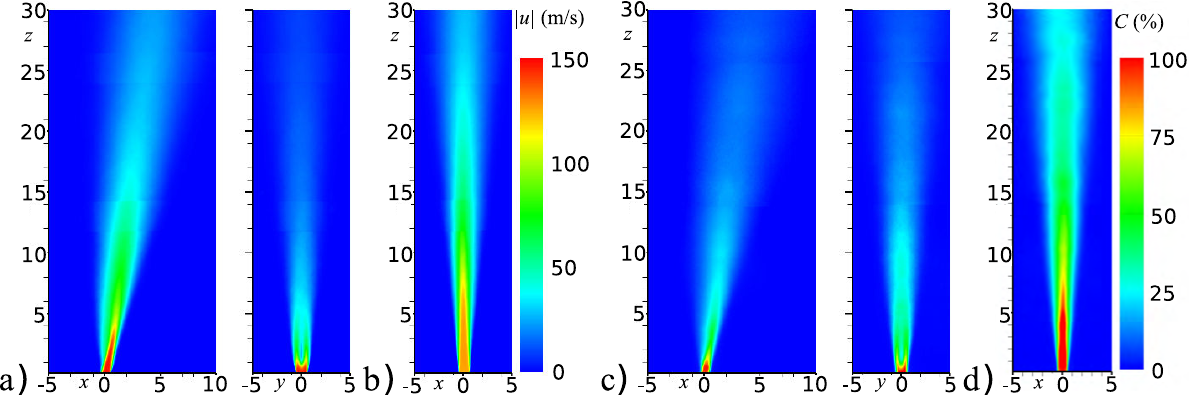}\\
	\raggedright \underline{\textbf{He:}}\\
	\includegraphics[scale=1.0]{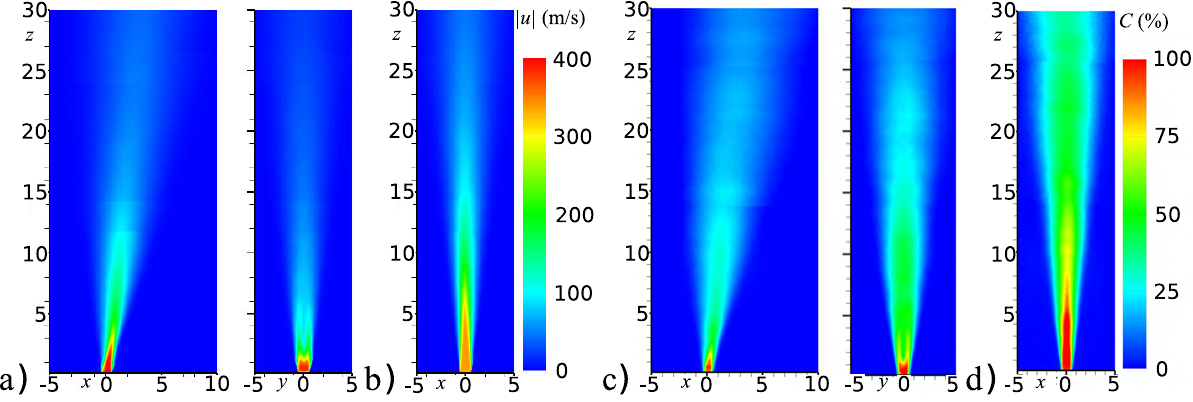}\\
	\caption{Time-averaged velocity and concentration contours in $x$-$z$ and $y$-$z$ planes for 1) air and 2) helium, obtained from a) Round jet on side of tube (3D jet) and b) Round orifice plate (OP) jet.}
	\label{fig.Velocity_Concentartion_Contours}
\end{figure}

The time-averaged velocity contours, obtained for all of the experiments, are shown in Fig.\ \ref{fig.Velocity_Concentartion_Contours}.  These contours are shown in both the $x$-$z$ and $y$-$z$ planes for the 3D jet experiments, and only {in} the $x$-$z$ plane for the OP jets. Clearly, for the 3D jets in the $x$-$z$ planes, there was a slight deviation from the vertical $z$-axis in the direction of the initial flow inside the pipe {for both gases.}  In this plane, significant jet spreading was observed as soon as the jets flowed through the orifice.  This {behaviour} was much more pronounced {in the case of the 3D jet} compared to the OP jet.  Also, for the 3D jets, near the potential-core region, there was more jet spreading on the back side of the jet (left side) compared to the front side. There was also a shorter potential-core length observed for helium compared to air.  These potential core lengths, in the $x$-$z$ planes, were approximately 4$D$ and 5$D$ for helium and air, respectively.  The potential-core lengths of both 3D jet gases were also shorter compared to the axisymmetric OP jets. The respective core lengths of the OP jets for helium and air were 7$D$ and 9$D$.  In the $x$-$z$ planes, for the 3D jet, the jet spreading appeared to be greater, in the far field, compared to the $y$-$z$ plane. There were also two high-velocity peaks observed in $y$-$z$ plane, for both gases, at $y\pm0.5D$, on each side of the $z$-axis, with a low-velocity region located on the axis at approximately $z=2D$.  These features were not observed in the OP jet.  Also, the potential-core lengths in this plane were much shorter compared to the $x$-$z$ plane.  In the $y$-$z$ plane, the potential-core length for both gases was approximately 1$D$.  In general, it was observed that the helium and air jets had qualitatively similar flow {patterns}, for each case. However, {the helium jet}, in both experiments, appeared to break up faster compared to {the air jet}.

Figure \ref{fig.Velocity_Concentartion_Contours} also shows the time-averaged concentration fields obtained for all experiments.  In general, the concentration profiles were qualitatively similar to the velocity profiles presented in Fig.\ \ref{fig.Velocity_Concentartion_Contours}, with two exceptions. First, much higher concentration levels, with higher spreading rates, were observed for helium in the far field compared to air, for all cases.  Also, for the 3D jets, the potential core lengths in the $x$-$z$ plane, for both gases, were comparable to the potential core lengths in the $y$-$z$ plane.

Although not shown here, numerically {computed} flow fields were also obtained for each gas.  It is noted that the fundamental asymmetry was also observed for each numerical experiment.

\subsection{The jet centreline trajectory}

Figure \ref{fig.centreline} shows the {experimental and numerical jet trajectories for the 3D jets.}  For the experiments, the trajectories were determined by the maximum velocity magnitude locations ($\lvert \boldsymbol{\overline{u}} \rvert_{\textrm{max}}(z)$), in the $x$-$z$ plane.  The numerical simulations present both the trajectories determined from ($\lvert \boldsymbol{\overline{u}} \rvert_{\textrm{max}}(z)$) and the centre of mass (C.M.).

\begin{figure}
	\centering
	\includegraphics[scale=0.333]{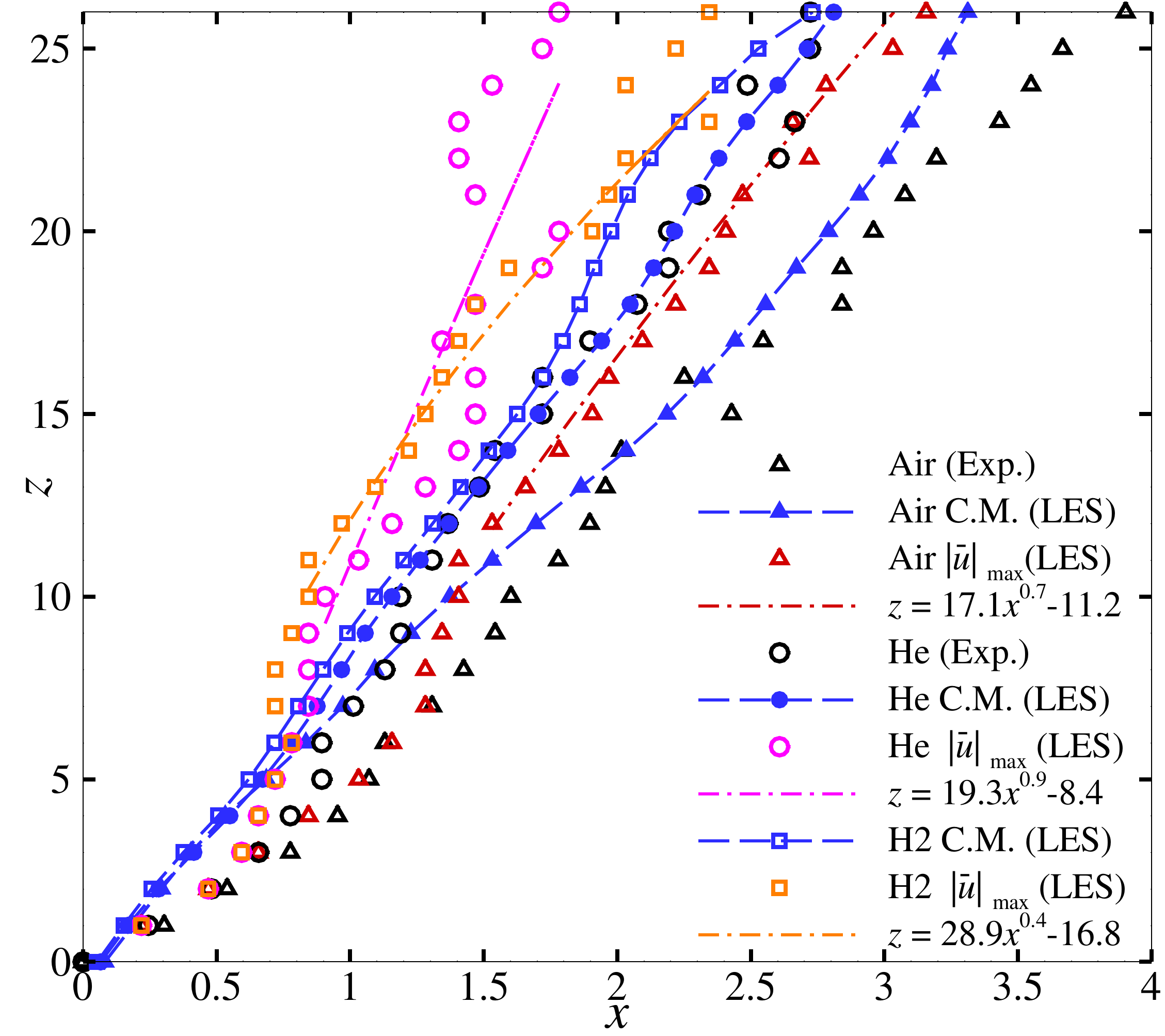}
	\caption{Jet centre-lines taken along the location of maximum velocity ($\lvert \boldsymbol{\overline{u}} \rvert_{\textrm{max}}$) locations from experiments and simulations, and also the centre of mass (C.M.) locations obtained from the simulations.}
	\label{fig.centreline}
\end{figure}

In general, the centrelines obtained from the experiments followed a nearly linear trajectory from the orifice up to $z\sim4D$ for helium, and $z\sim5D$ for air.  From these point, the centrelines deviated upwards, slightly.  Around $z\sim9D$ for helium, and $z\sim10D$ for air, a sudden change in the jet trajectory was observed.  These locations coincided with the extents of the potential-core regions as {shown} in Fig.\ \ref{fig.Velocity_Concentartion_Contours}.  {Also, helium was found to deviate from the $z$-axis less than air, owing to increased buoyancy forces, in the $z$-direction, which arise due to the lower jet gas density.}

From the simulations, it was clear that the jet centres, determined from $\lvert \boldsymbol{\overline{u}} \rvert_{\textrm{max}}(z)$, matched well the experimental observations in the near field.  However, in the far field, {the} simulated locations do not match those obtained from the experiments for helium and air beyond $z>5D$ and $z>7D$, respectively.  Despite this, however, it was found that the C.M.\ locations matched very well the jet centre-lines obtained from the experiments in the far field, beyond these locations, for both helium and air.  Clearly, the simulations exhibited a departure of the $\lvert \boldsymbol{\overline{u}} \rvert_{\textrm{max}}(z)$ location from the actual jet centroid through the entire jet height for all three gases.  Also, the $\lvert \boldsymbol{\overline{u}} \rvert_{\textrm{max}}(z)$ locations for the helium simulation were found to contain significant scatter in the far field beyond $z>15D$.  The air and hydrogen simulations, on the other hand, were found to have a fairly continuous trajectory determined from the $\lvert \boldsymbol{\overline{u}} \rvert_{\textrm{max}}$ locations.  As a reference, lines of best-fit, using linear regression to power-law expressions, were obtained for the far field (beyond $z>10D$) for the centrelines determined from $\lvert \boldsymbol{\overline{u}} \rvert_{\textrm{max}}(z)$.  Even though helium deflected more than hydrogen, in terms of the C.M.\ locations, the opposite trend was observed when considering the $\lvert \boldsymbol{\overline{u}} \rvert_{\textrm{max}}(z)$ locations.

\subsection{Velocity decay and jet spreading rates}

In Fig.\ \ref{fig.centrelineVelocity}a, the velocity decay along the jet centrelines, determined from $\lvert \boldsymbol{\overline{u}} \rvert_{\textrm{max}}(z)$, are presented for {all cases}, both experimental and numerical.  Also shown, for comparison, are velocity decay correlations \cite{Witze1974AJ417}, which have been determined from {prior} compressible subsonic and supersonic axisymmetric round jet experiments, for the range jet conditions that encompass the current investigation.  Upon comparison to the Witze correlations \cite{Witze1974AJ417}, the air and helium OP jet experiments were found to reproduce well the expected velocity decay rate, with helium decaying faster than the air jet.  In general, the decay rates observed in the experimental 3D jets were much faster compared to the axisymmetric jets.  Upon comparison of the experimental 3D jet velocity decays to simulation, it was found that the simulated air jet velocity decay rate matched closely that obtained from experiment.  For the helium jet, however, the simulation had a much faster decay rate compared to experiment.  Despite this, both exhibit the same trend, where helium was found to decay faster than air. It was also observed that the experiments had a shorter potential-core region compared to simulation.  In general, {the simulated onsets of velocity decay, downstream from the orifice, were found to occur approximately 2$D$ beyond those observed from the experiments.}  Finally, from simulation, hydrogen {jet} was found to decay the quickest, owing to its low density and high flow velocity.

\begin{figure}
	\centering
	a)\includegraphics[scale=0.245]{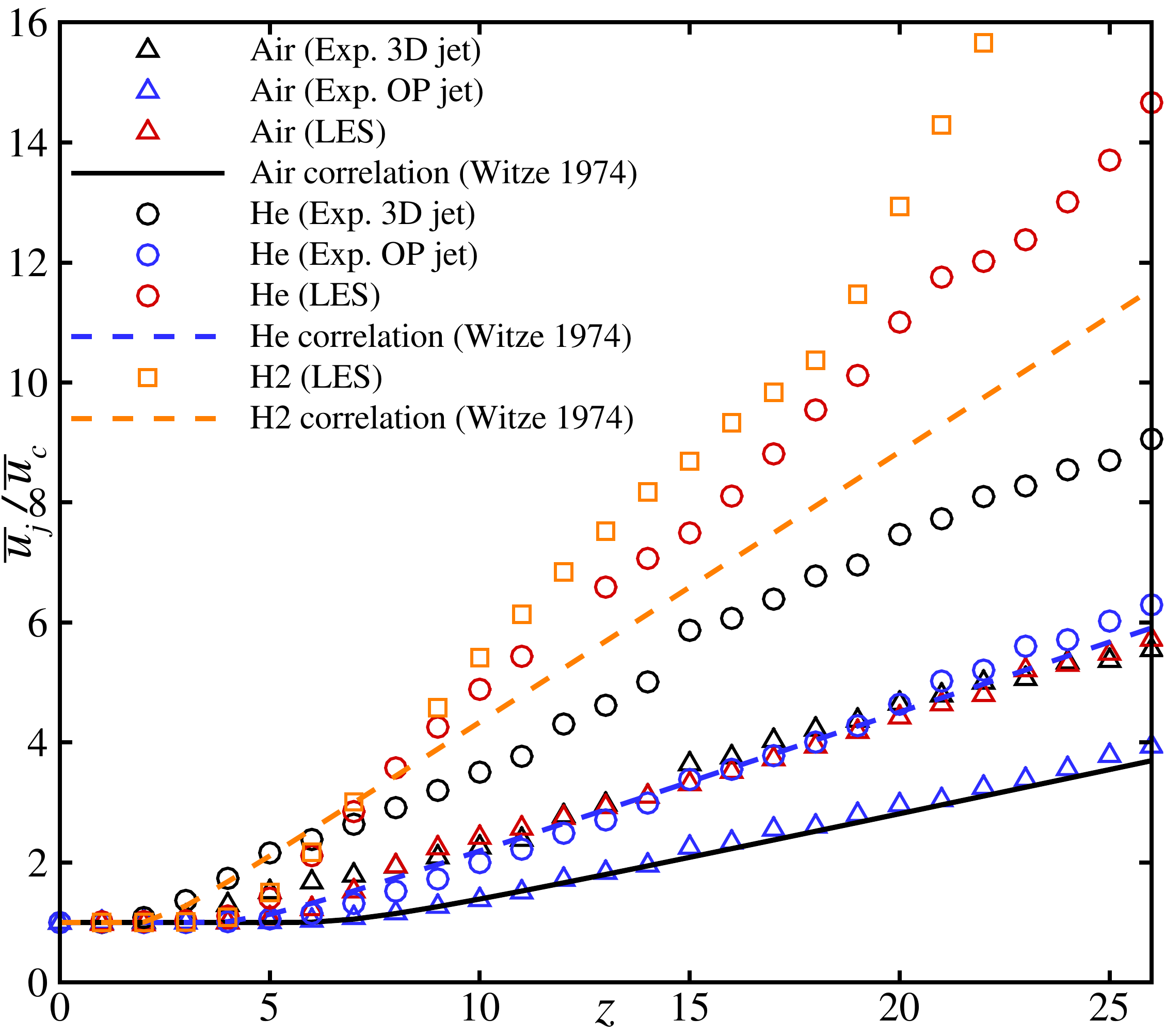} 
	b)\includegraphics[scale=0.245]{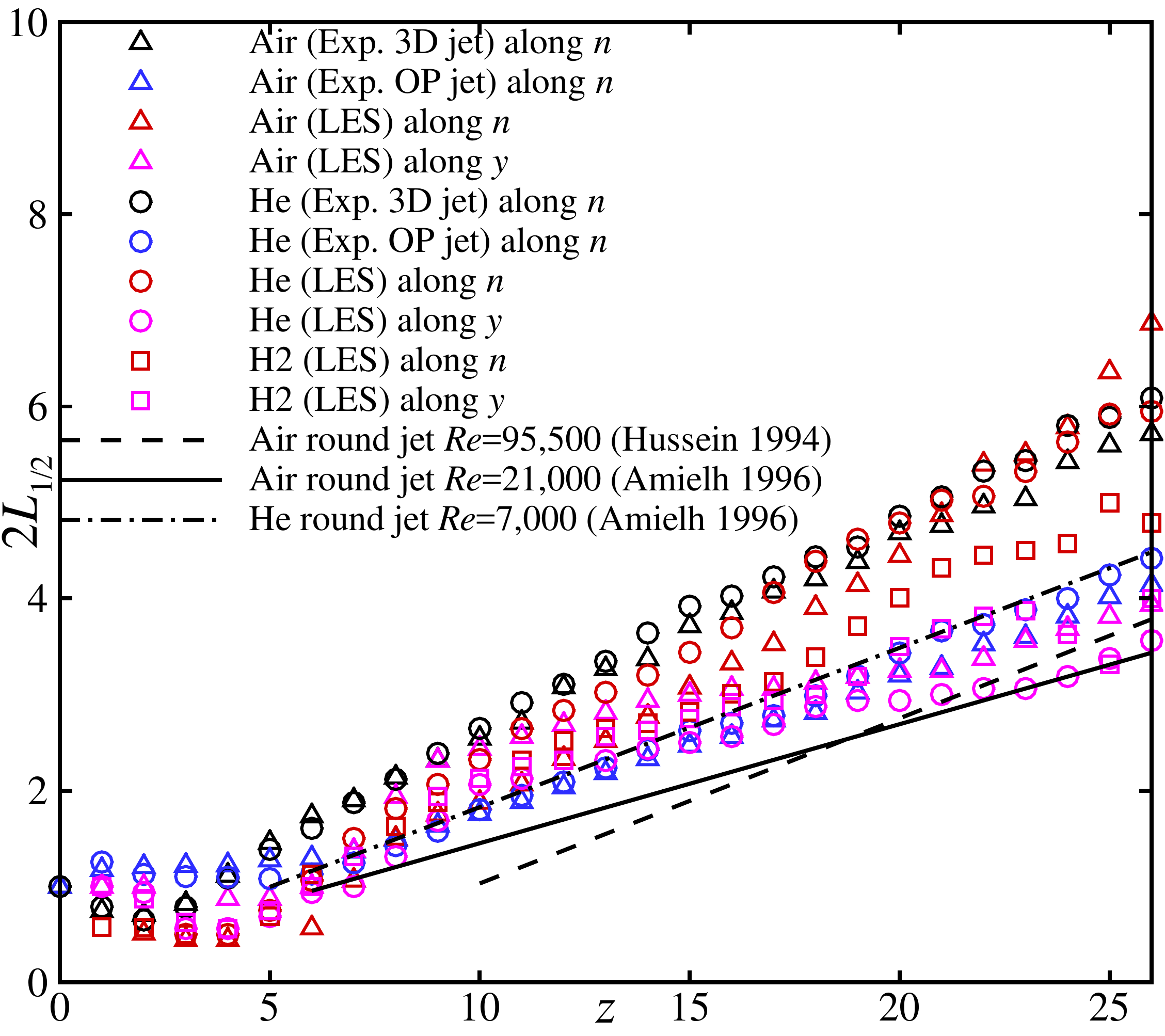}
	\caption{a) Jet velocity decay and b) jet widths (2$L_{1/2}$) obtained along the $\lvert \boldsymbol{\overline{u}} \rvert_{\textrm{max}}$ centrelines, from simulations and experiments.  Note, the $n$-coordinate refers to lines which are normal to the centreline, coplanar with the $x$-$z$ plane (see the coordinate system in Fig.\ref{fig.Experimental_Layout} b).  Also, velocity decays and jet widths have been compared to axisymmetric round jet correlations \cite{Witze1974AJ417} and experiments \cite{Hussein1994JoFM31,Amielh1996IJoHaMT2149}, respectively.}
	\label{fig.centrelineVelocity}
\end{figure}

Figure \ref{fig.centrelineVelocity}b {shows} the jet widths (2$L_{1/2}$) {that} have been obtained by determining the locations where $\lvert\boldsymbol{\overline{u}}\rvert=0.5\lvert \boldsymbol{\overline{u}} \rvert_{\textrm{max}}(z)$ along lines which were orthogonal to the jet-centrelines.  These orthogonal lines have been indicated previously as coordinate $n$ in Fig.\ \ref{fig.Experimental_Layout}b.  In the $y$-$z$ planes, the orthogonal lines to the jet-centres are collinear with the $y$-direction owing to symmetry of the jet.  Also, jet widths along the $y$-direction from the jet centreline were only available from simulation owing to the three-dimensional deflection of the jet centre in the $x$-direction.

In general, the OP jets were found to have nearly constant jet widths in the potential core region, up until $z\sim5$.  From this point on, the jet width was found to increase linearly, consistent with the jet spreading rates of previous axisymmetric round jet experiments (for  a wide range of $Re$) \cite{Hussein1994JoFM31,Amielh1996IJoHaMT2149}. For the 3D jets, in all cases, a slight contraction in the jet widths has been observed from $1<z<4$ experimentally, and from $1<z<7$ numerically.  Beyond this point, the jet spreading rates in the $x$-$z$ plane, along $n$, were observed to be much greater compared to the axisymmetic jets for all cases.  Moreover, the air and helium jet spreading, from the 3D experiments and also the simulations, was found to be comparable for both gases.  In the near field, however, significant jet spreading {did} not occur until about $z\sim7$ to 8 for the simulations, while spreading was found to occur at $z\sim5$ for the 3D experiments.  These values coincide with the potential-core extents of the jets.  In the $y$-$z$ plane, along $y$, the jet spreading obtained from the simulations deviated from those obtained in the $x$-$z$ plane for $z>12D$.  In fact, it was found that the jet spreading of air and helium in the $y$-direction, found numerically, compared well to the jet spreading of axisymmetric round jets, in terms of order of magnitude, while more jet spreading was found in the $x$-$z$ plane.  Finally, unlike air and helium, the simulated hydrogen 3D jet was found to spread almost equally in both directions.  There was only slightly higher amount of jet spreading in the $n$-direction compared to the $y$-direction.

\subsection{Jet centreline statistics}

In the $x$-$z$ plane, the time-averaged velocity profiles for all experimental and numerical investigations are shown in Fig. \ref{fig.Vs} along the $n$-direction for several downstream locations along the jet centreline ($s$-curve Fig.\ \ref{fig.Experimental_Layout}b).  Likewise, the velocity profiles are also shown along the $y$-direction.  It is noted, however, that information along $y$, normal to $s$, was only available from {the} numerical simulation.  Also shown, for comparison, are the velocity statistics obtained for the OP jet experiments.  In particular, only the $s$-component velocities have been presented, which were normalized by the local centreline velocity magnitudes ($\lvert \boldsymbol{\overline{u}} \rvert_{\textrm{c}}(z)$).  Also, the $n$- and $y$-coordinates, which were both normal to the centreline $s$-curve, were normalized by the jet half widths ($L_{1/2}$) determined from Fig.\ref{fig.centrelineVelocity}b.  In all cases, the experimental and numerical 3D jets emerged from the orifice with a top-hat profile, shown at $z=1$.  This was different compared to the OP jet, which had an initial semi saddle-back profile, typical for axisymmetric sharp-edged OP jets  \cite{Mi2007EiF625}.  In general, all cases of the experimental and numerical 3D jets developed into a self-similar Gaussian-like distribution within the range $\lvert n/L_{1/2}\rvert < 1$ {for $z\le5$} (and $\lvert y/L_{1/2}\rvert < 1$ {for $z\le5$}).  In fact, the distribution observed for the 3D jets, in this range, matched well the self-similar Gaussian distribution obtained from the OP jets.  However, notable deviations from the self-similar solution were observed near the tail ends of the curves in the $x$-$z$ plane, beyond this range .  The experiments were found to exhibit more velocity spreading to the left of the jet centre (in the $-n$-direction).  On the other hand, the simulations were found to exhibit more velocity spreading to the right of the jet centre (in the $+n$-direction).  Beyond {$z>5D$}, in the far field, the experimental 3D jets developed into, and matched, the self-similar Gaussian distribution obtained from the OP jets for the full range of $n$ (and $y$).  The numerical simulations, however, continued to exhibit velocity spreading to the right of the jet centre (in the $+n$-direction).  Finally, the curves obtained for all three gases were found to be in agreement with each other.

\begin{figure} 
	\centering
	\raggedright \underline{\textbf{air:}}\\
	\centering
	a)\includegraphics[scale=0.25,trim={0.1cm 0.1cm 0.1cm 0.1cm},clip]{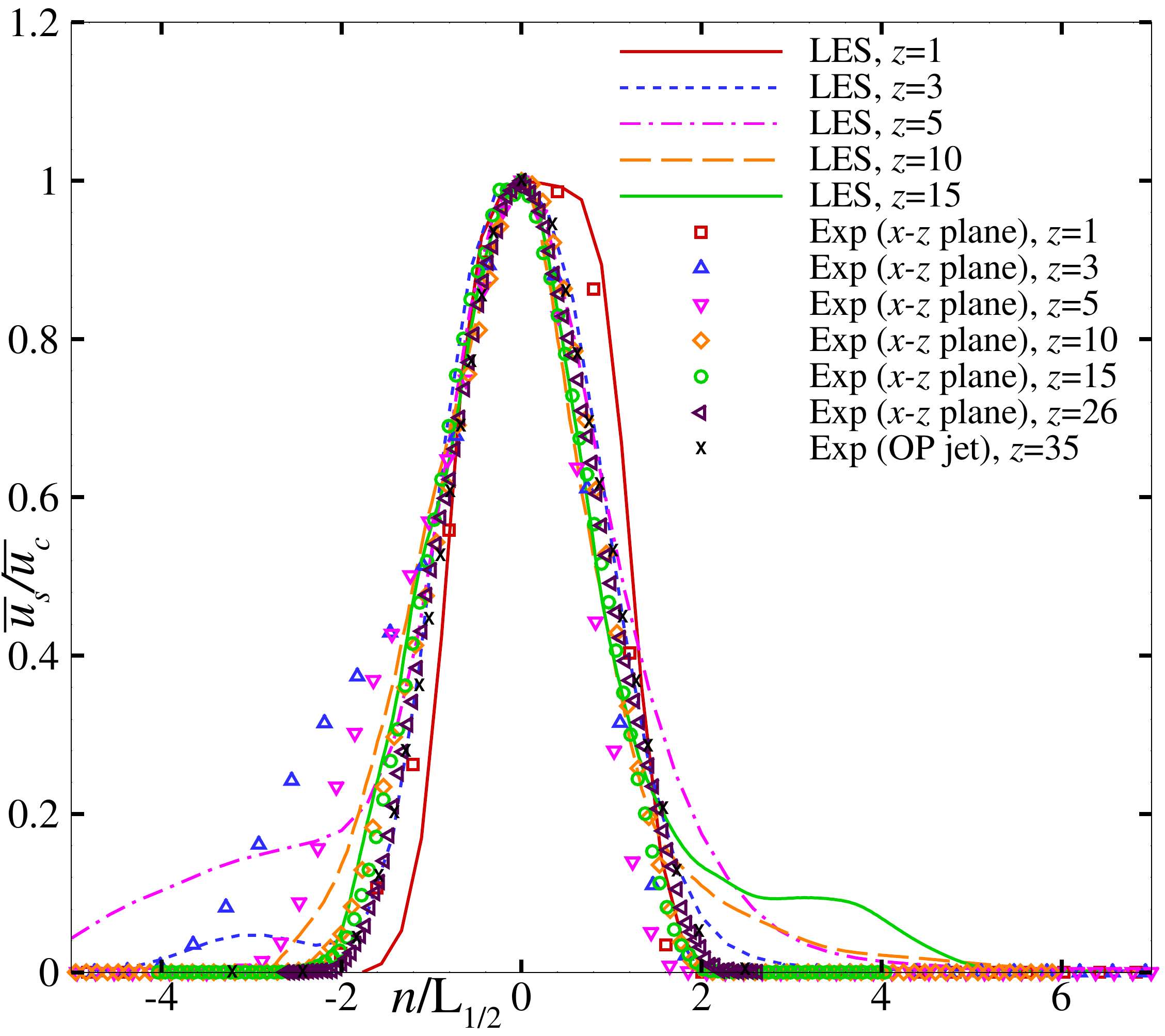}
	b)\includegraphics[scale=0.25,trim={0.1cm 0.1cm 0.1cm 0.1cm},clip]{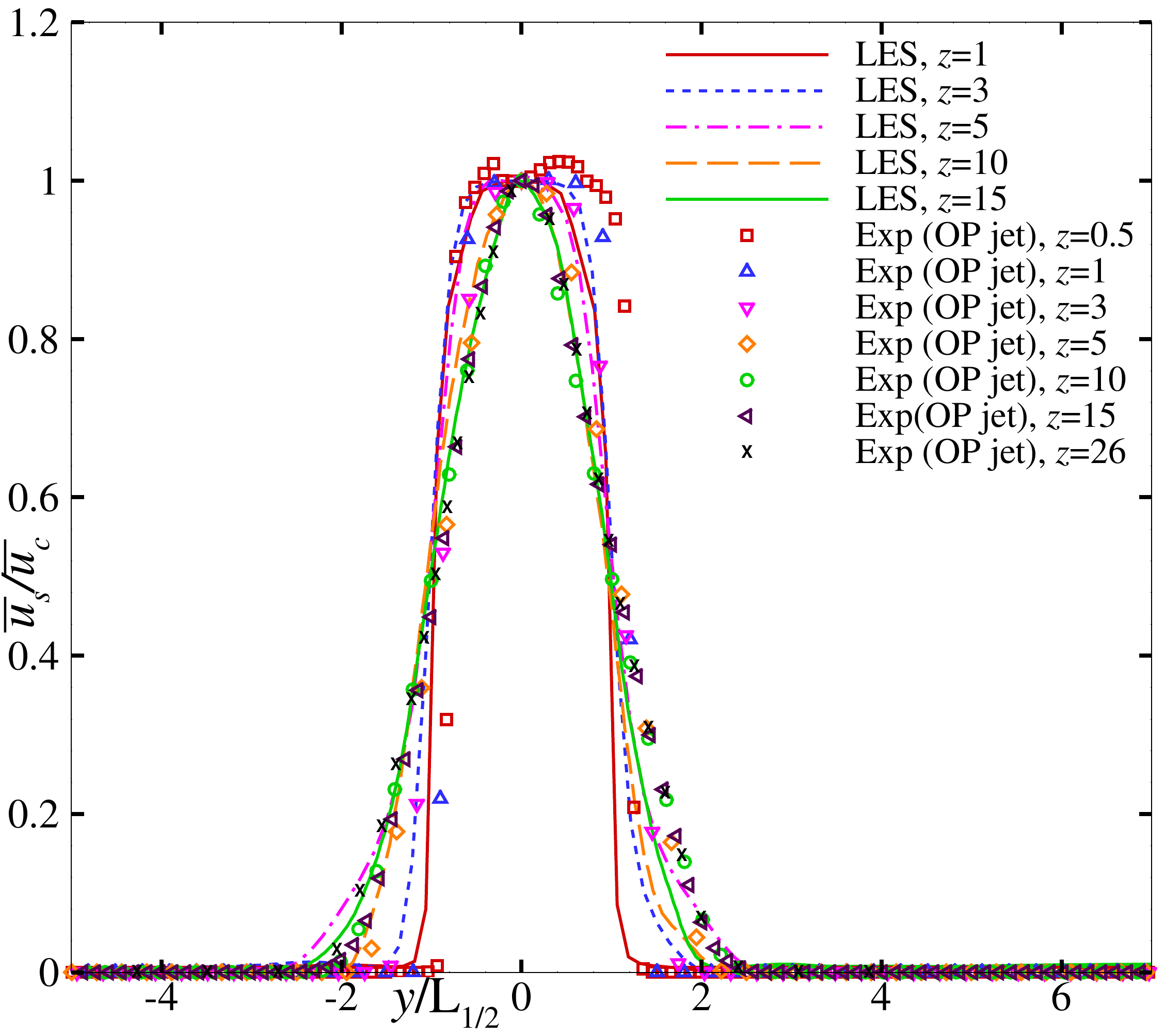}\\
	\raggedright \underline{\textbf{He:}}\\
	\centering
	a)\includegraphics[scale=0.25,trim={0.1cm 0.1cm 0.1cm 0.1cm},clip]{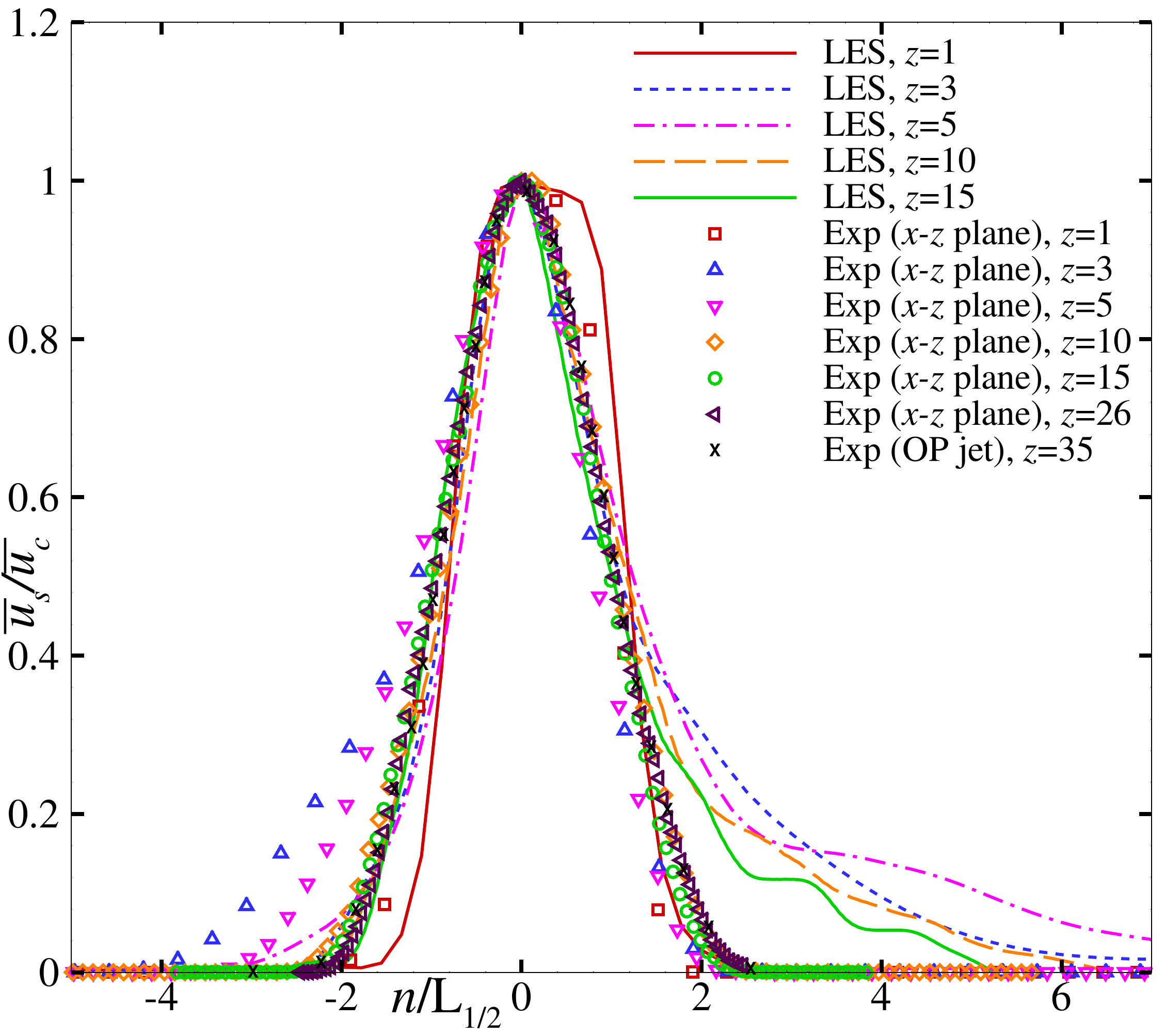}
	b)\includegraphics[scale=0.25,trim={0.1cm 0.1cm 0.1cm 0.1cm},clip]{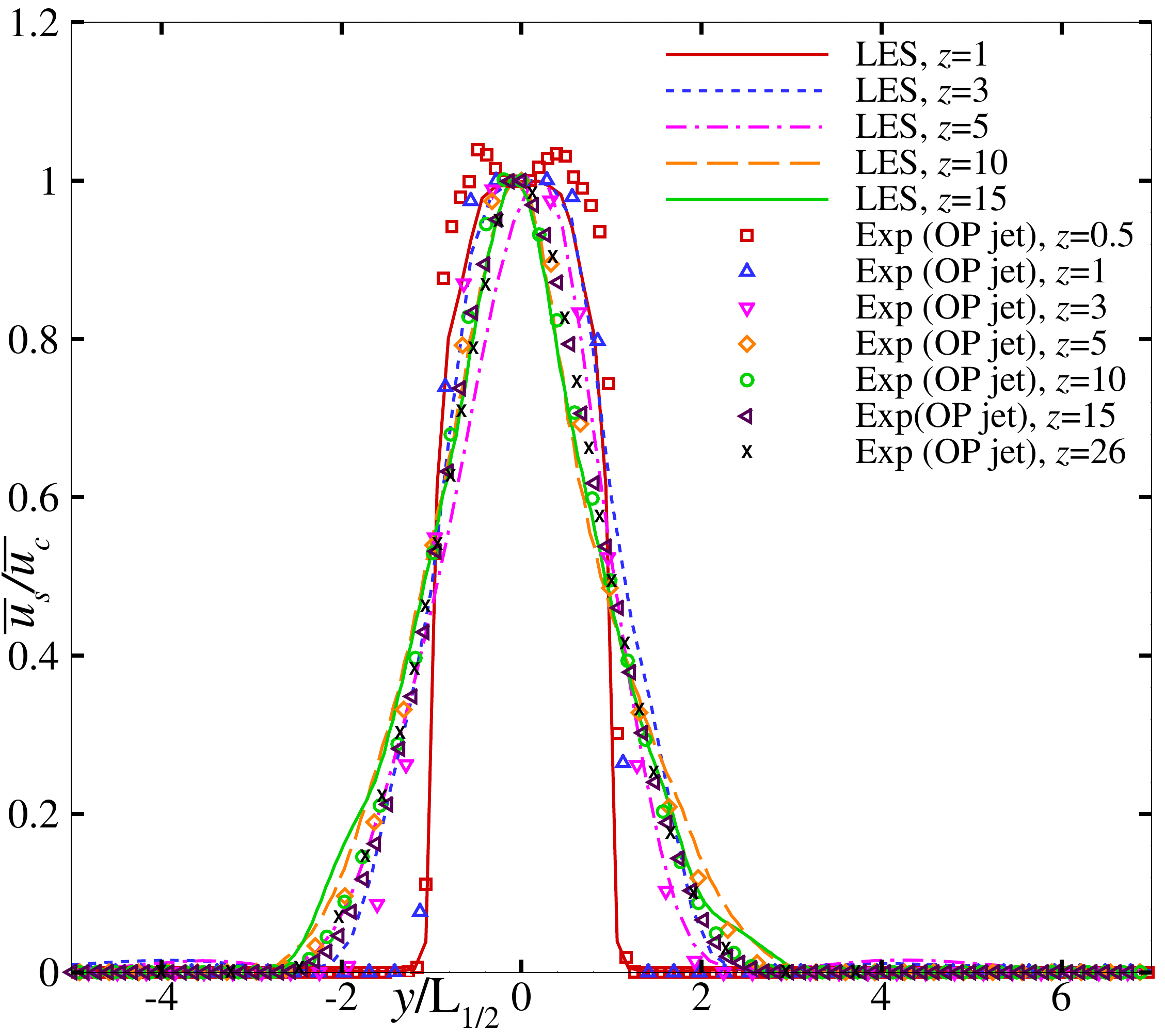}\\
	\raggedright \underline{\textbf{$\textrm{H}_2$:}}\\
	\centering
	a)\includegraphics[scale=0.25,trim={0.1cm 0.1cm 0.1cm 0.1cm},clip]{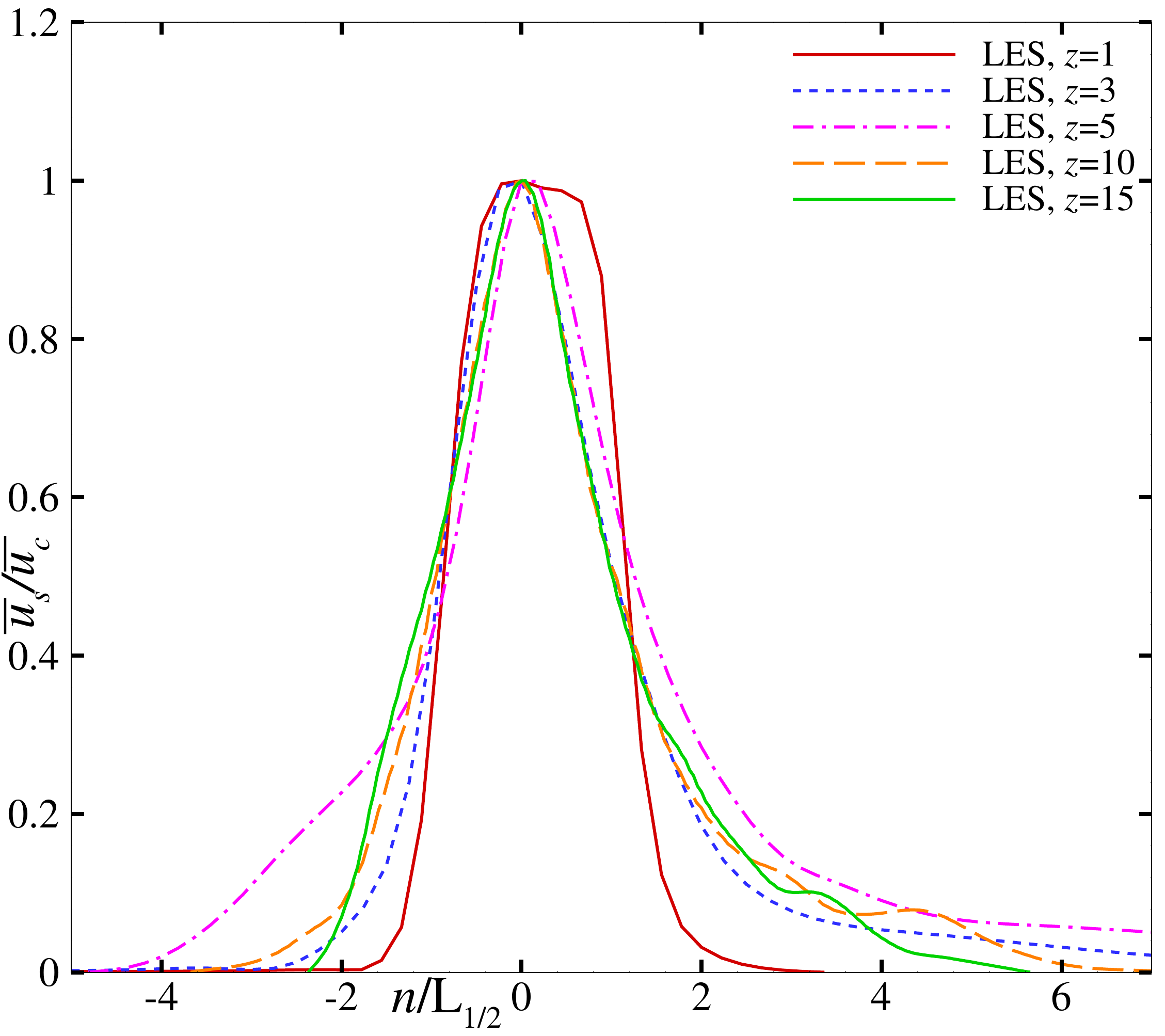}
	b)\includegraphics[scale=0.25,trim={0.1cm 0.1cm 0.1cm 0.1cm},clip]{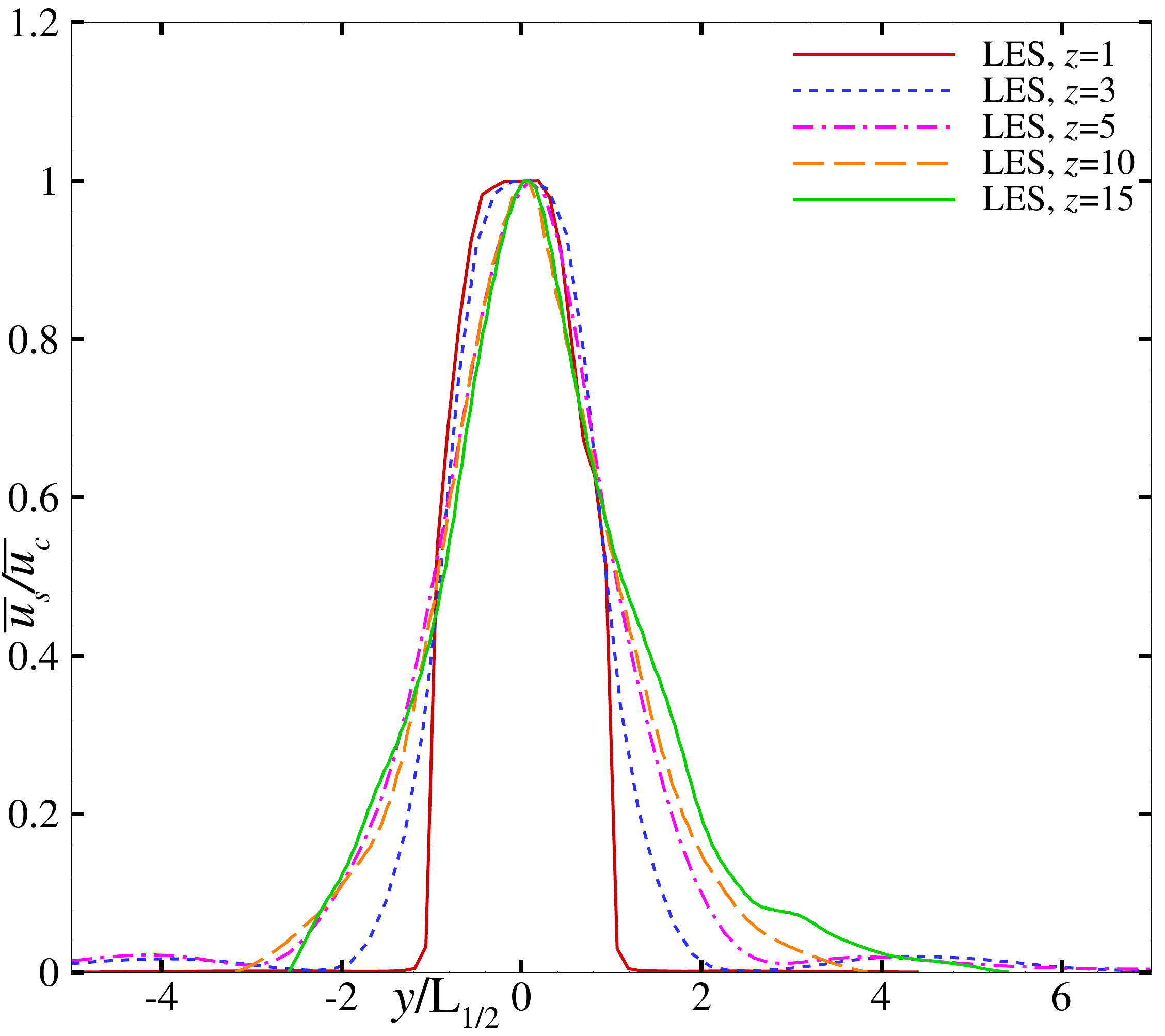}\\
	\caption{{Time-averaged} velocity profiles along jet centrelines, taken at various heights for air, helium, and hydrogen, {and} obtained from a) LES, OP \& 3D jet in $x$-$z$ plane and b) LES \& OP jet in $y$-$z$ planes.}
	\label{fig.Vs}
\end{figure}

Higher order statistics were also collected for the experiments and simulations conducted here.  {Time-}averaged Reynolds stress profiles ($\overline{u_s'u_n'}$ and $\overline{u_s'u_y'}$) {are} presented in Fig.\ \ref{fig.Reynolds_Stresses}, following the same format at the velocity profiles in Fig.\ \ref{fig.Vs} but normalized by $\lvert \boldsymbol{\overline{u}} \rvert^2_{\textrm{c}}(z)$.  In the $x$-$z$ plane, the air and helium experiments captured well the far field self-similar profile to the right of the jet centre (in the $+n$-direction).  However, to the left (in the $-n$-direction), the 3D jet experiments were found to have a lower magnitude of {the} Reynolds stress compared to the OP jets, even in the far field.  Also, within for $z\le5$, a higher reynolds stress was observed beyond $\lvert n/L_{1/2}\rvert < 1$.  In the $y$-$z$ plane, the 3D jet experiments matched well the Reynolds stress profiles obtained from the OP jets.  In general, the simulations for air and helium captured the correct shapes of the Reynolds stress profiles, but did not capture the peak values observed experimentally, in both planes.  This {effect can be} attributed to the fact that the Reynolds stress was determined from $\overline{\tilde{u}_s'\tilde{u}_y'}$ and $\overline{\tilde{u}_s'\tilde{u}_n'}$, which did not account for the subgrid contribution from $k_{sgs}$.

    \begin{figure} 
    	\centering
    	\raggedright \underline{\textbf{air:}}\\
    	\centering
    	a)\includegraphics[scale=0.25,trim={0.1cm 0.1cm 0.1cm 0.1cm},clip]{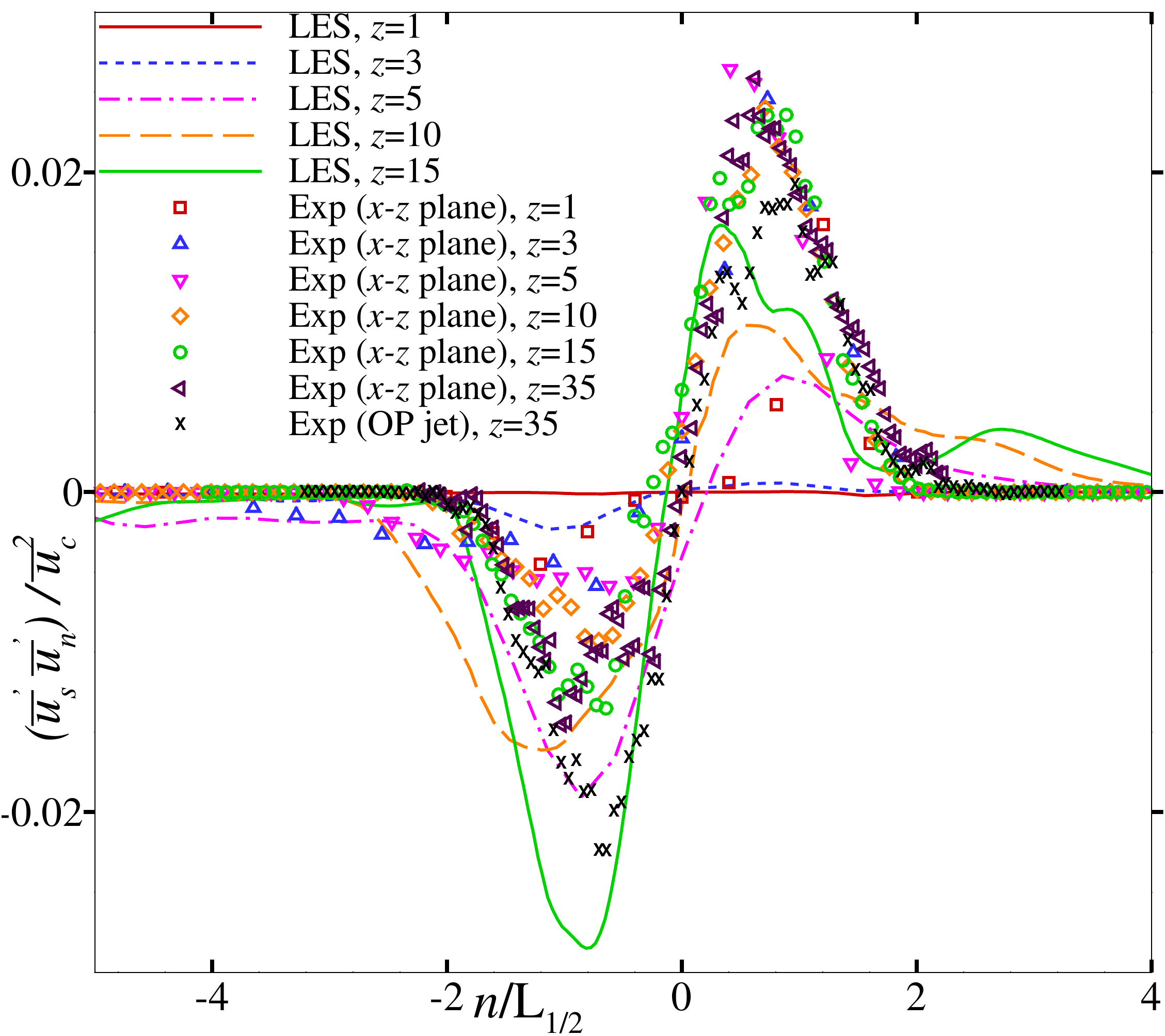}
    	b)\includegraphics[scale=0.25,trim={0.1cm 0.1cm 0.1cm 0.1cm},clip]{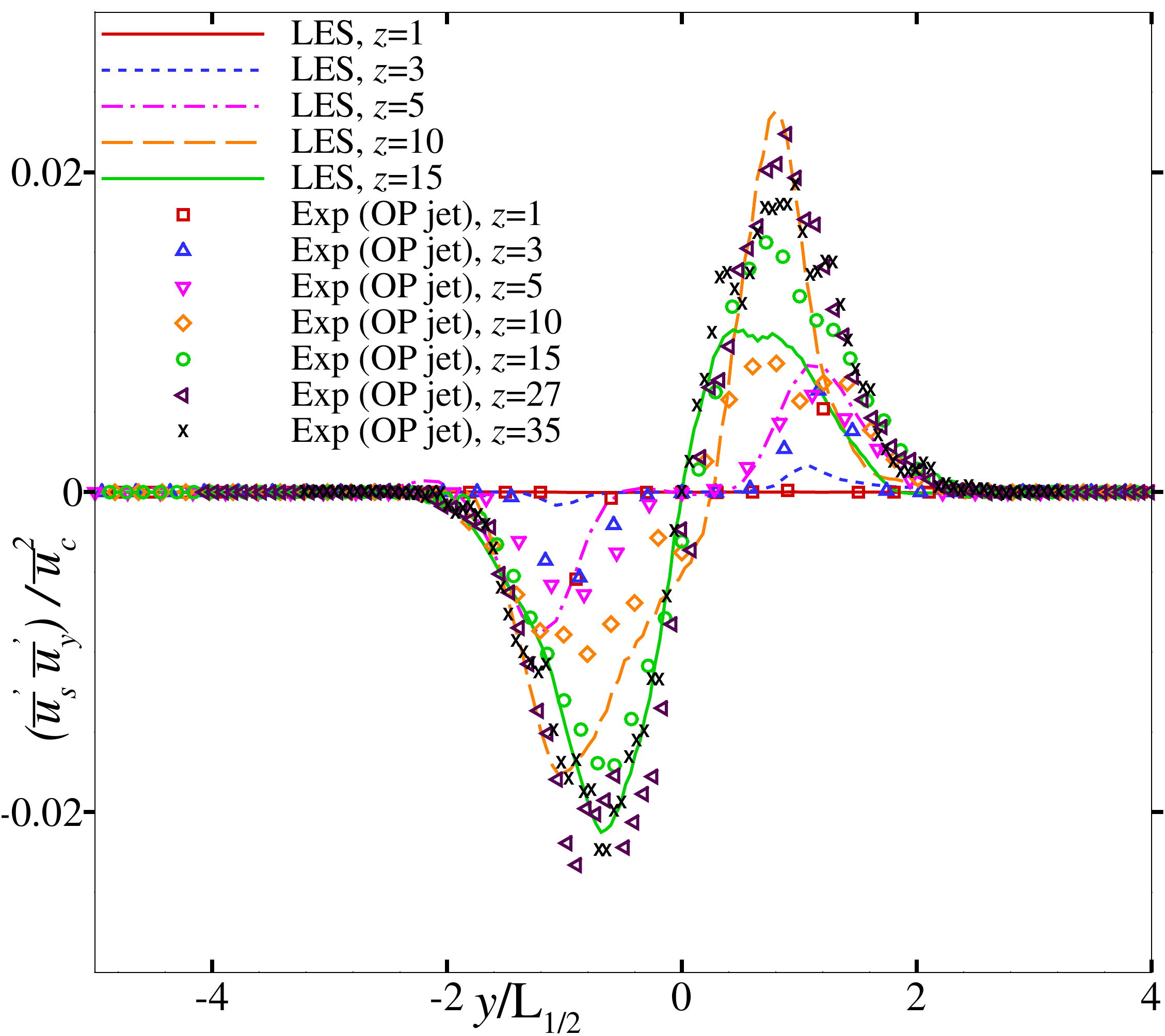}\\
    	\raggedright \underline{\textbf{He:}}\\
    	\centering
    	a)\includegraphics[scale=0.25,trim={0.1cm 0.1cm 0.1cm 0.1cm},clip]{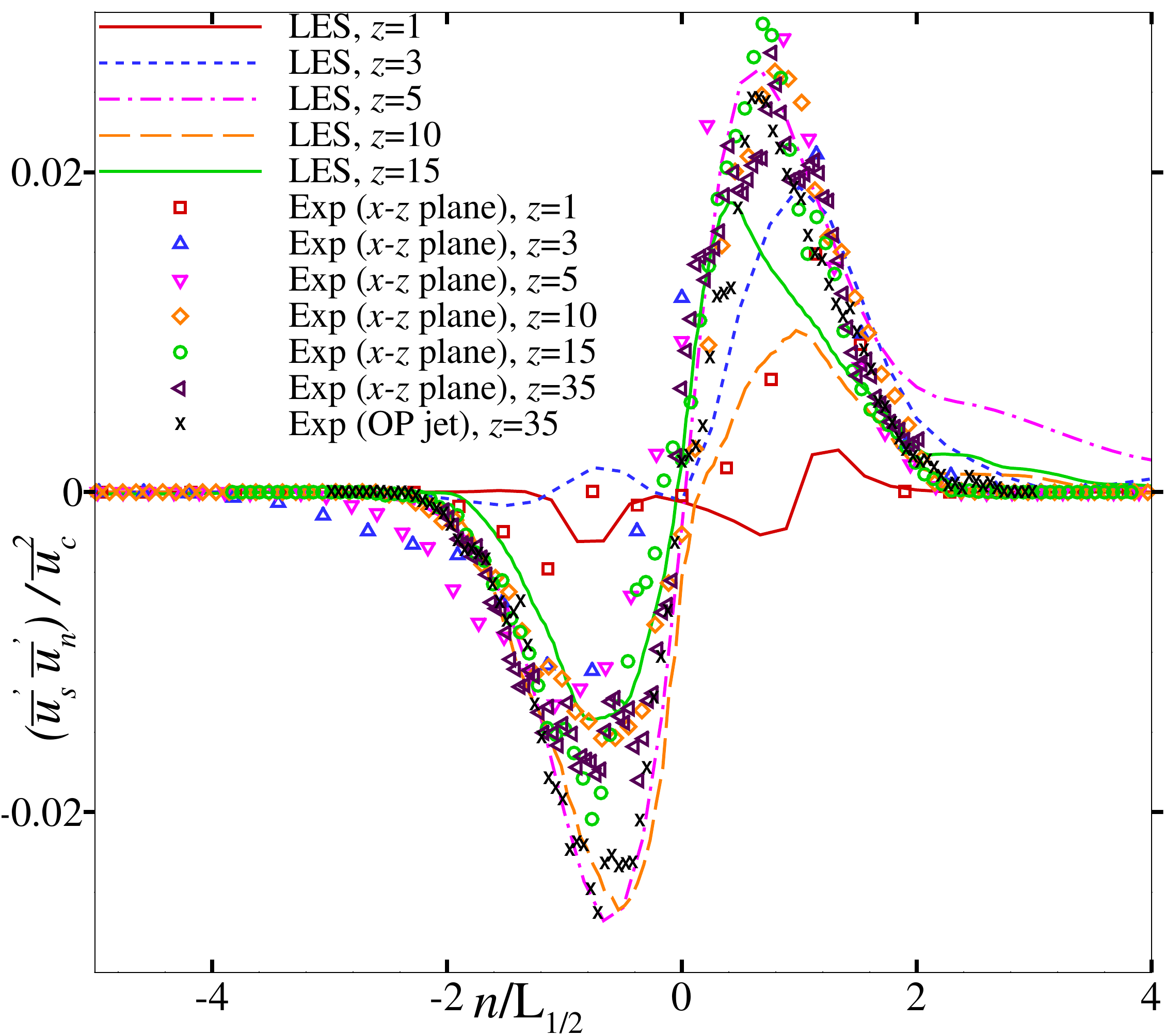}
    	b)\includegraphics[scale=0.25,trim={0.1cm 0.1cm 0.1cm 0.1cm},clip]{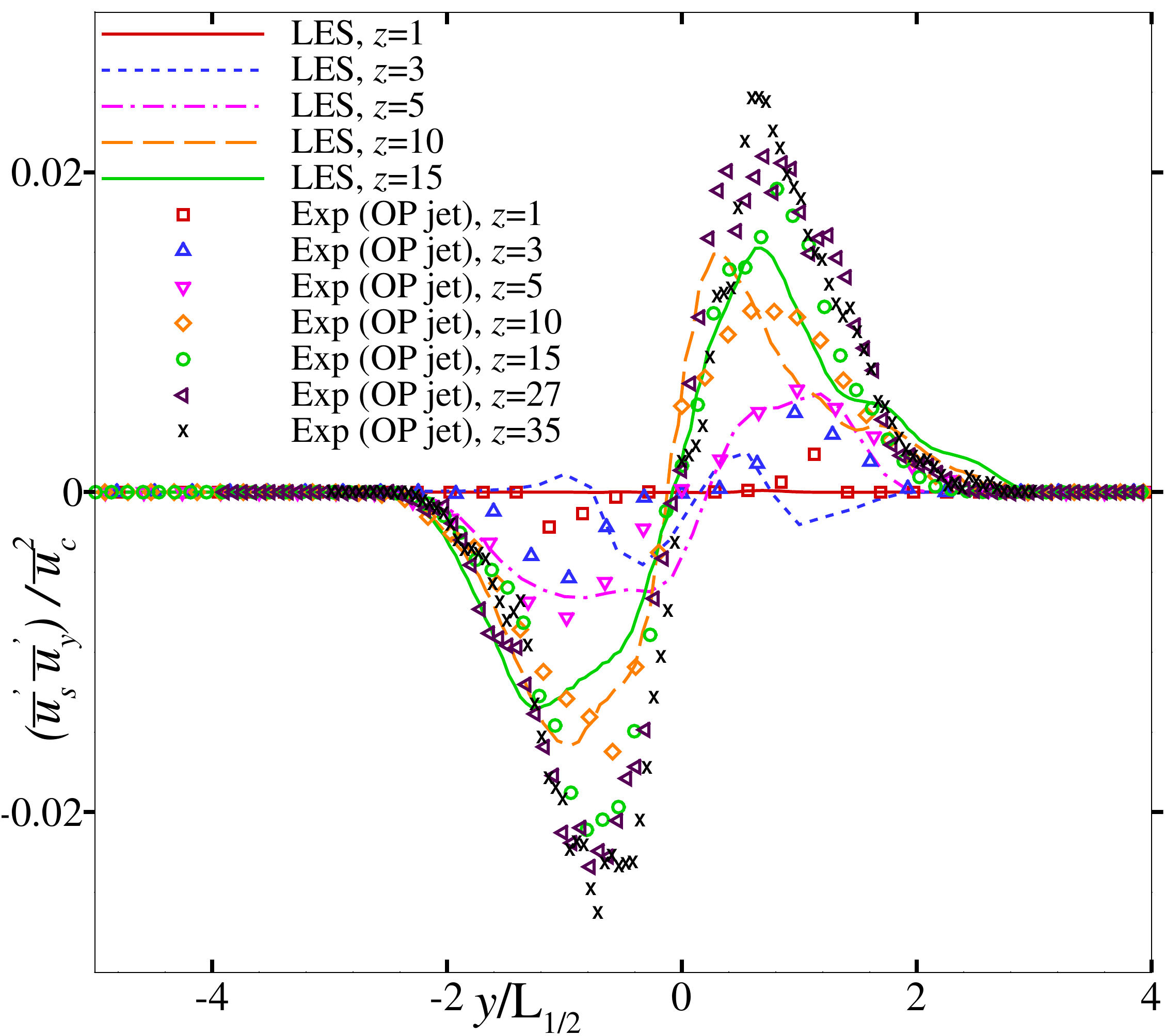}\\
    	\raggedright \underline{\textbf{$\textrm{H}_2$:}}\\
    	\centering
    	a)\includegraphics[scale=0.25,trim={0.1cm 0.1cm 0.1cm 0.1cm},clip]{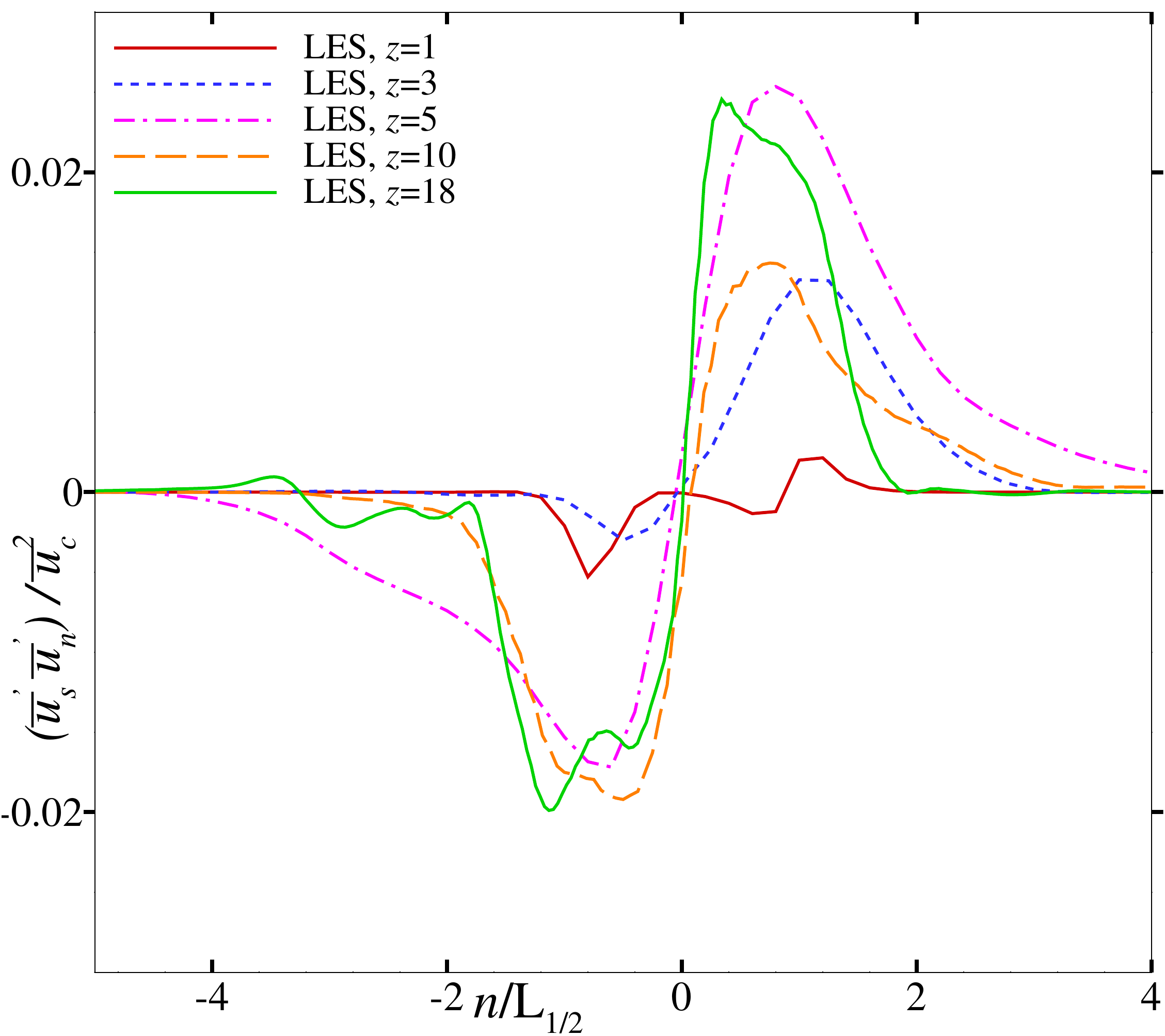}
    	b)\includegraphics[scale=0.25,trim={0.1cm 0.1cm 0.1cm 0.1cm},clip]{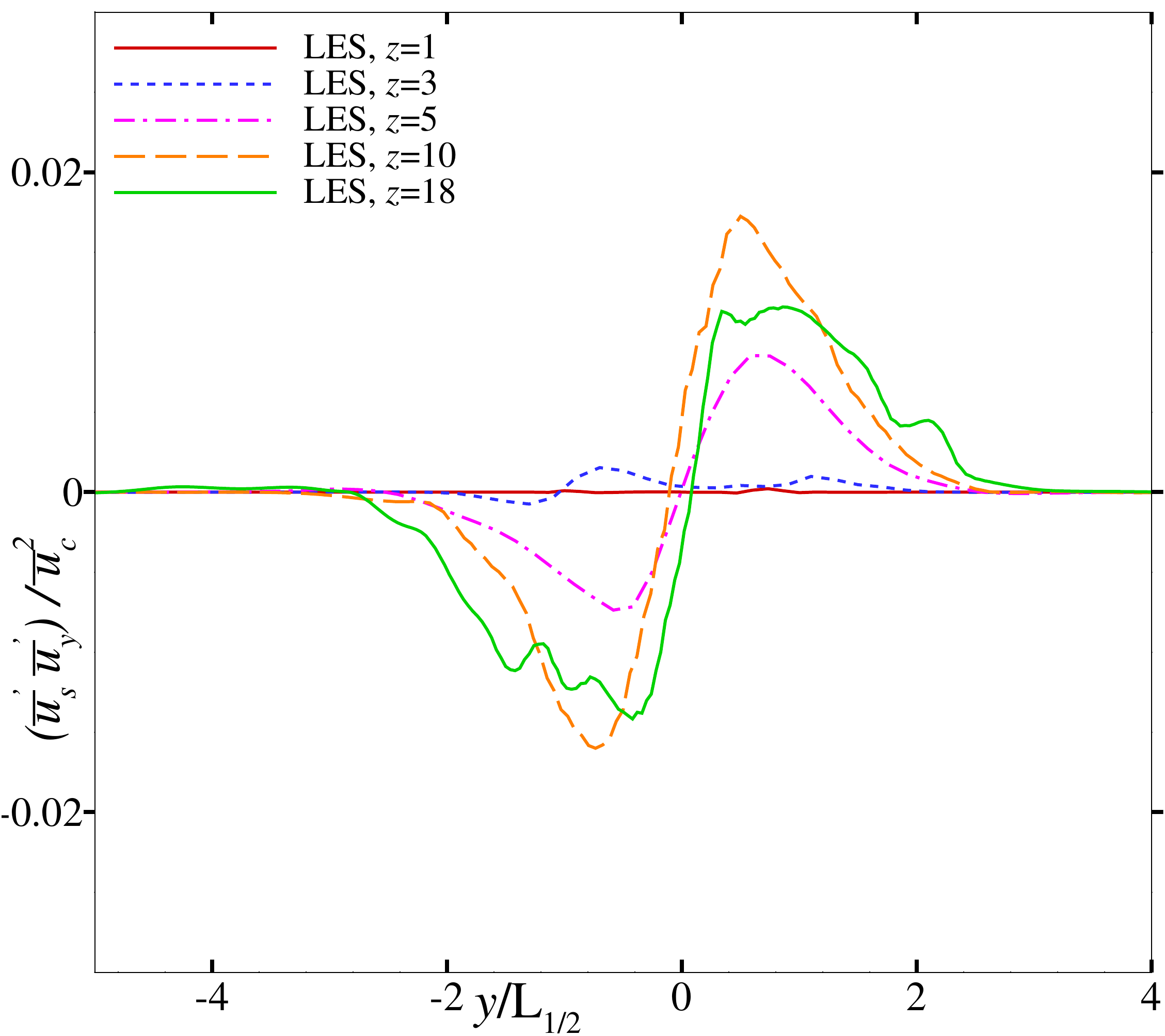}\\
    	\caption{{Time-averaged} Reynolds shear stress profiles along jet centrelines, taken at various heights for  air, helium, and hydrogen, {and} obtained from a) LES, OP \& 3D jet in $x$-$z$ plane and b) LES \& OP jet in $y$-$z$ planes.}
    	\label{fig.Reynolds_Stresses}
    \end{figure}

Finally, the time-averaged concentration profiles for all experimental and numerical cases are shown in Fig.\ \ref{fig.Cs}. Here, molar concentrations ($\overline{C}$), have been normalized by the local centreline concentrations ($\overline{C_c}(z)$). The $n$- and $y$-coordinates were normalized by the jet half widths ($L_{c,1/2}$) determined from the locations where $\overline{C/C_c}=0.5$.  In general, they were found to be {qualitatively} similar to the velocity profiles in all cases.  In the near field ($z\le5D$), the range $\lvert n/L_{c,1/2}\rvert < 1$ {for $z\le5D$} (and $\lvert y/L_{c,1/2}\rvert < 1$ {for $z\le5D$}) was found to develop quickly into the self-similar solution as observed from the OP jet experiments.  Notable deviations from the self-similar solution were once again observed near the tail ends of the curves in the $x$-$z$ plane, beyond this range.  In the $x$-$z$plane, the experiments were found to exhibit more concentration spreading to the left of the jet centre (in the $-n$-direction), while the simulations were found to exhibit more concentration spreading to the right (in the $+n$-direction).  In the far field, beyond $z>5D$, the self-similar Gaussian distribution, as observed for the OP jet experiments, was recovered for both the 3D jet experiments.  As observed in the velocity profiles, the simulations continued to exhibit concentration spreading in the $+n$-direction of the far field.

\begin{figure} 
	\centering
	\raggedright \underline{\textbf{air:}}\\
	\centering
	a)\includegraphics[scale=0.25,trim={0.1cm 0.1cm 0.1cm 0.1cm},clip]{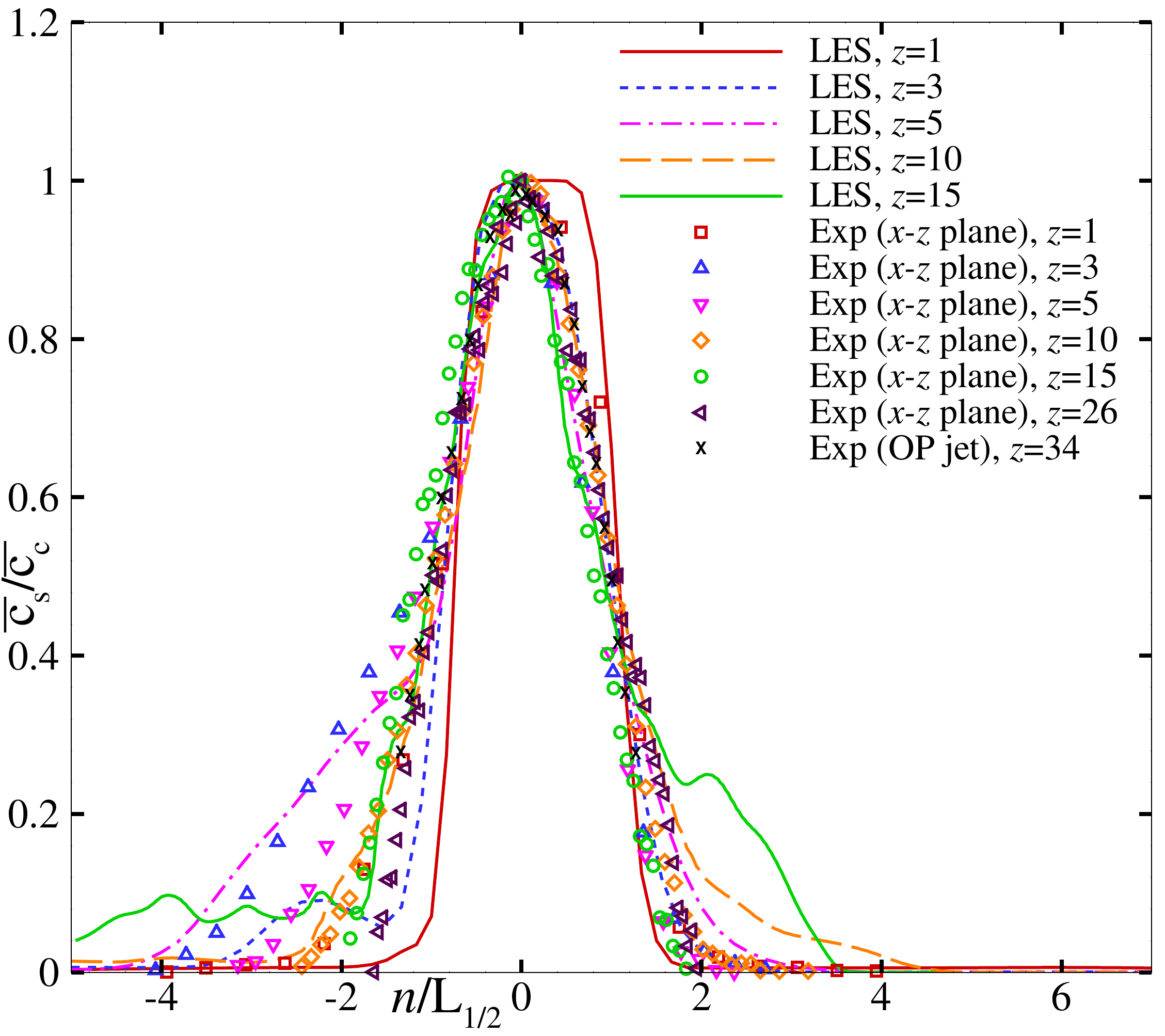}
	b)\includegraphics[scale=0.25,trim={0.1cm 0.1cm 0.1cm 0.1cm},clip]{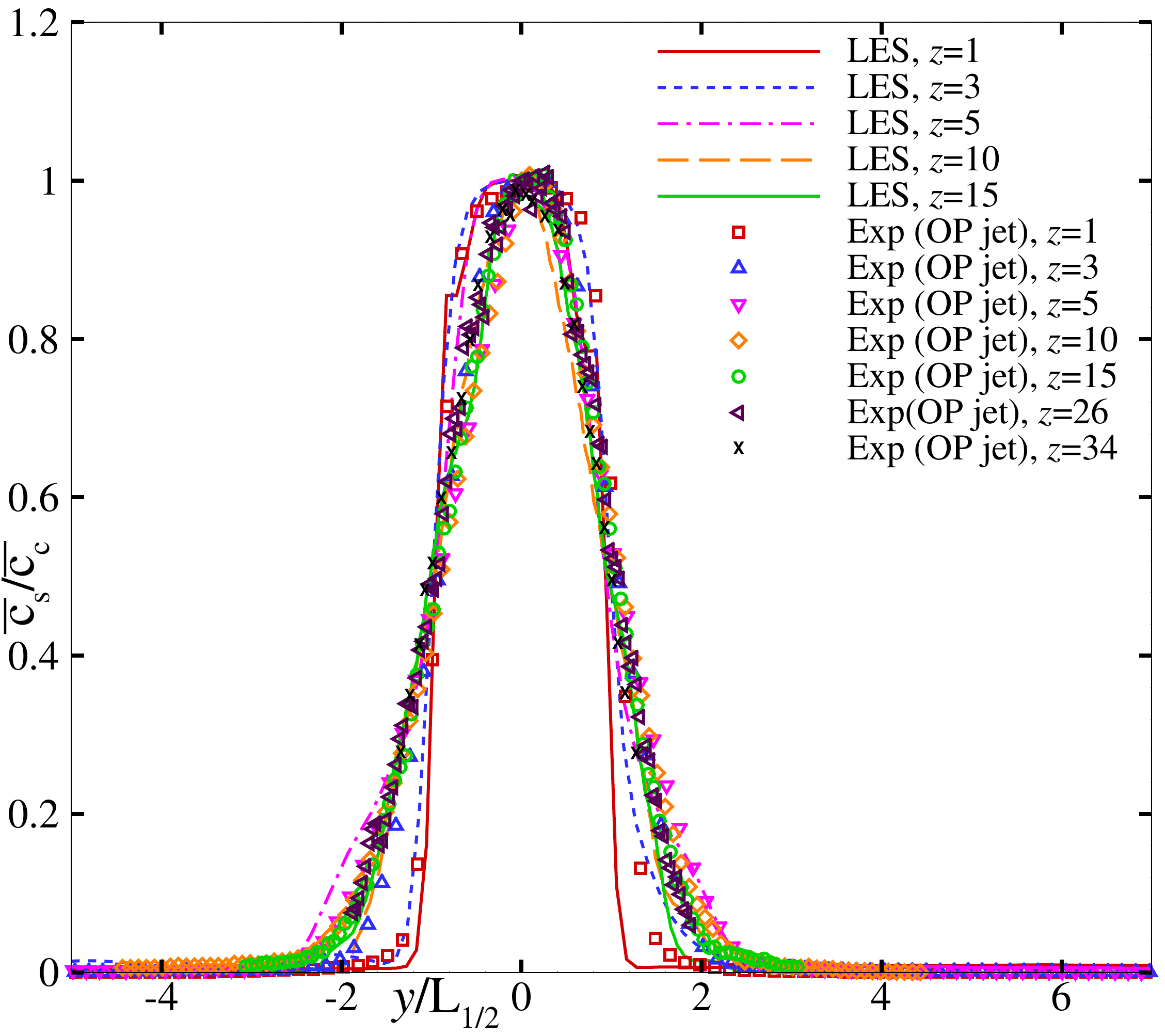}\\
	\raggedright \underline{\textbf{He:}}\\
	\centering
	a)\includegraphics[scale=0.25,trim={0.1cm 0.1cm 0.1cm 0.1cm},clip]{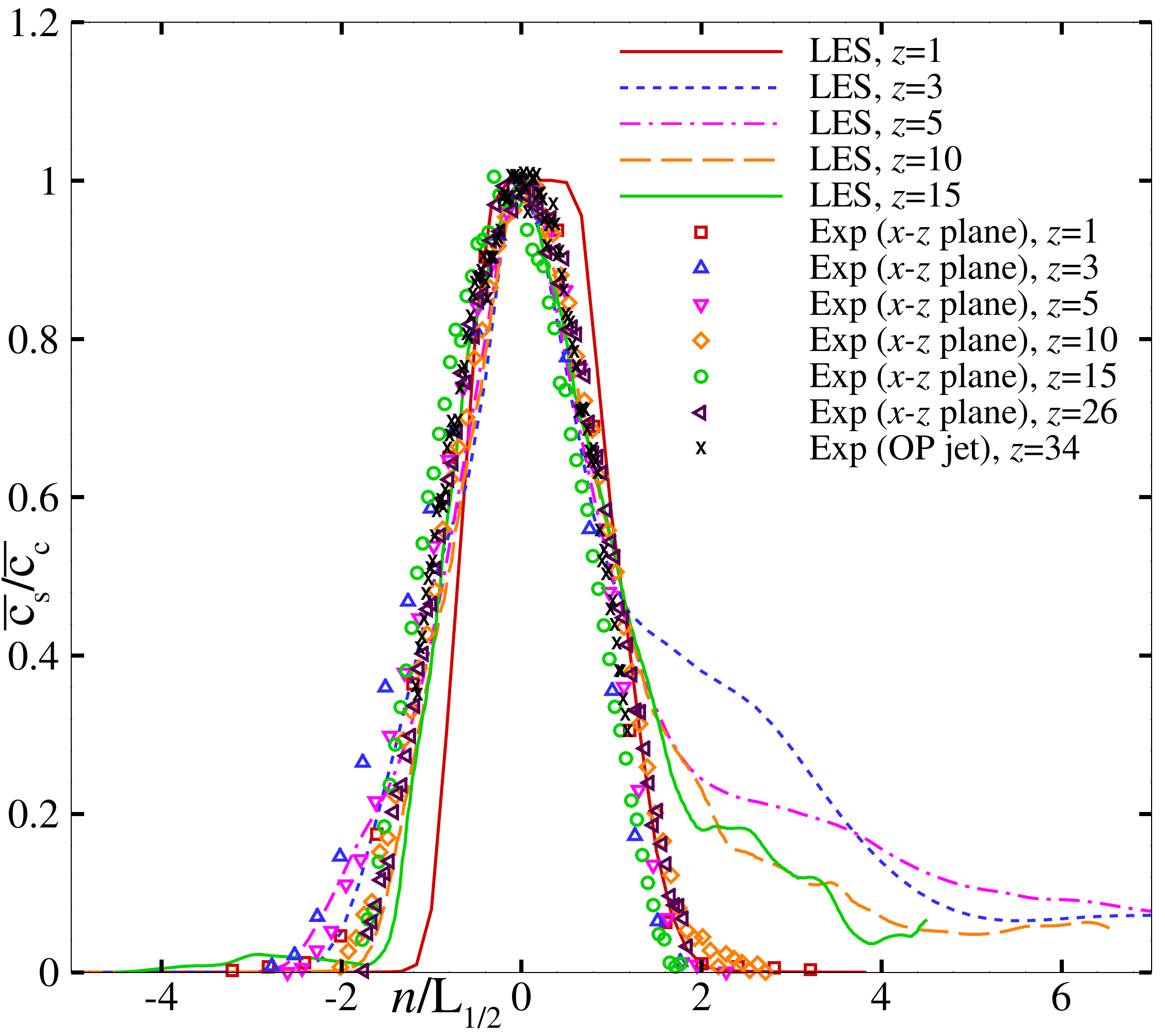}
	b)\includegraphics[scale=0.25,trim={0.1cm 0.1cm 0.1cm 0.1cm},clip]{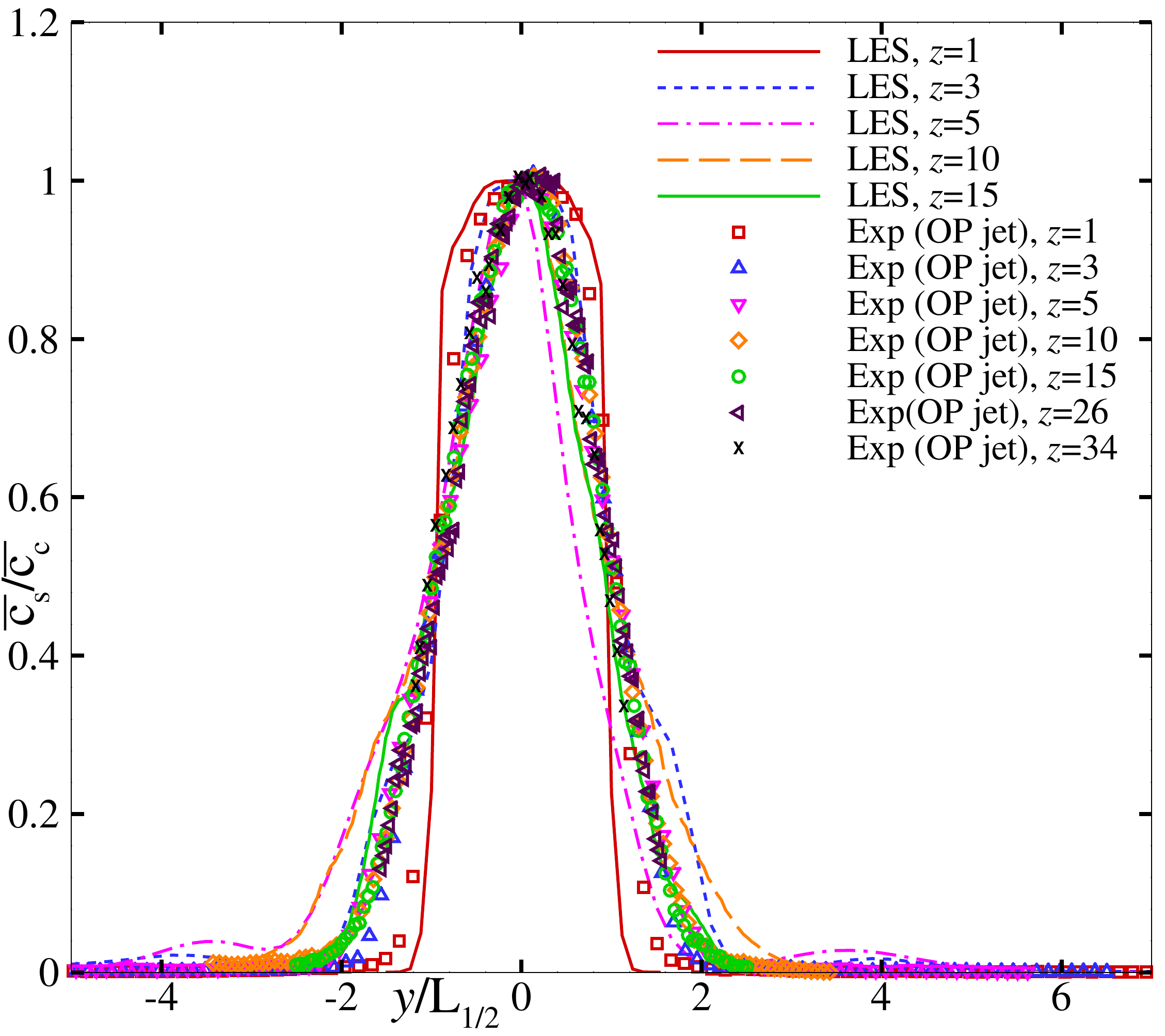}\\
	\raggedright \underline{\textbf{$\textrm{H}_2$:}}\\
	\centering
	a)\includegraphics[scale=0.25,trim={0.1cm 0.1cm 0.1cm 0.1cm},clip]{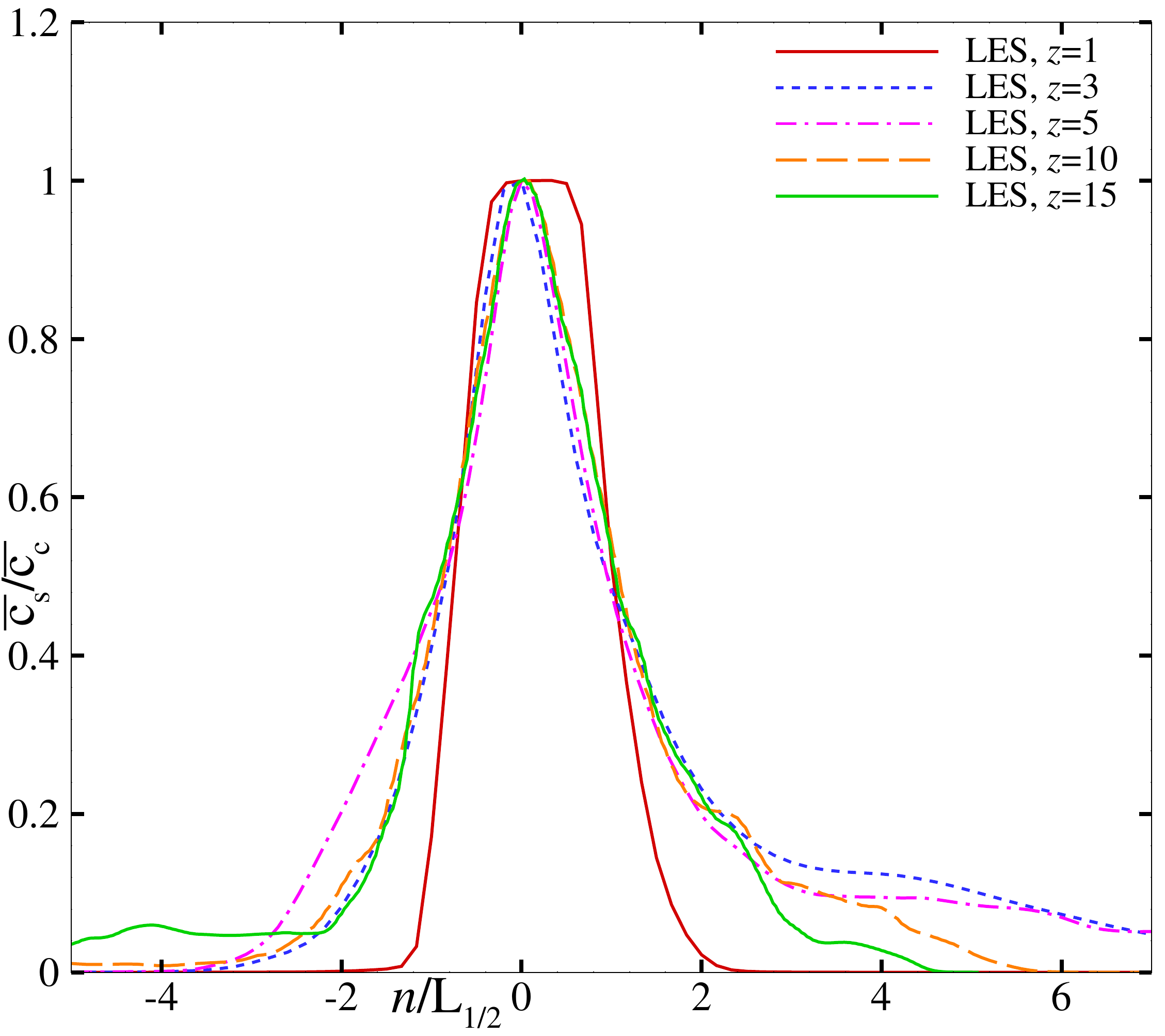}
	b)\includegraphics[scale=0.25,trim={0.1cm 0.1cm 0.1cm 0.1cm},clip]{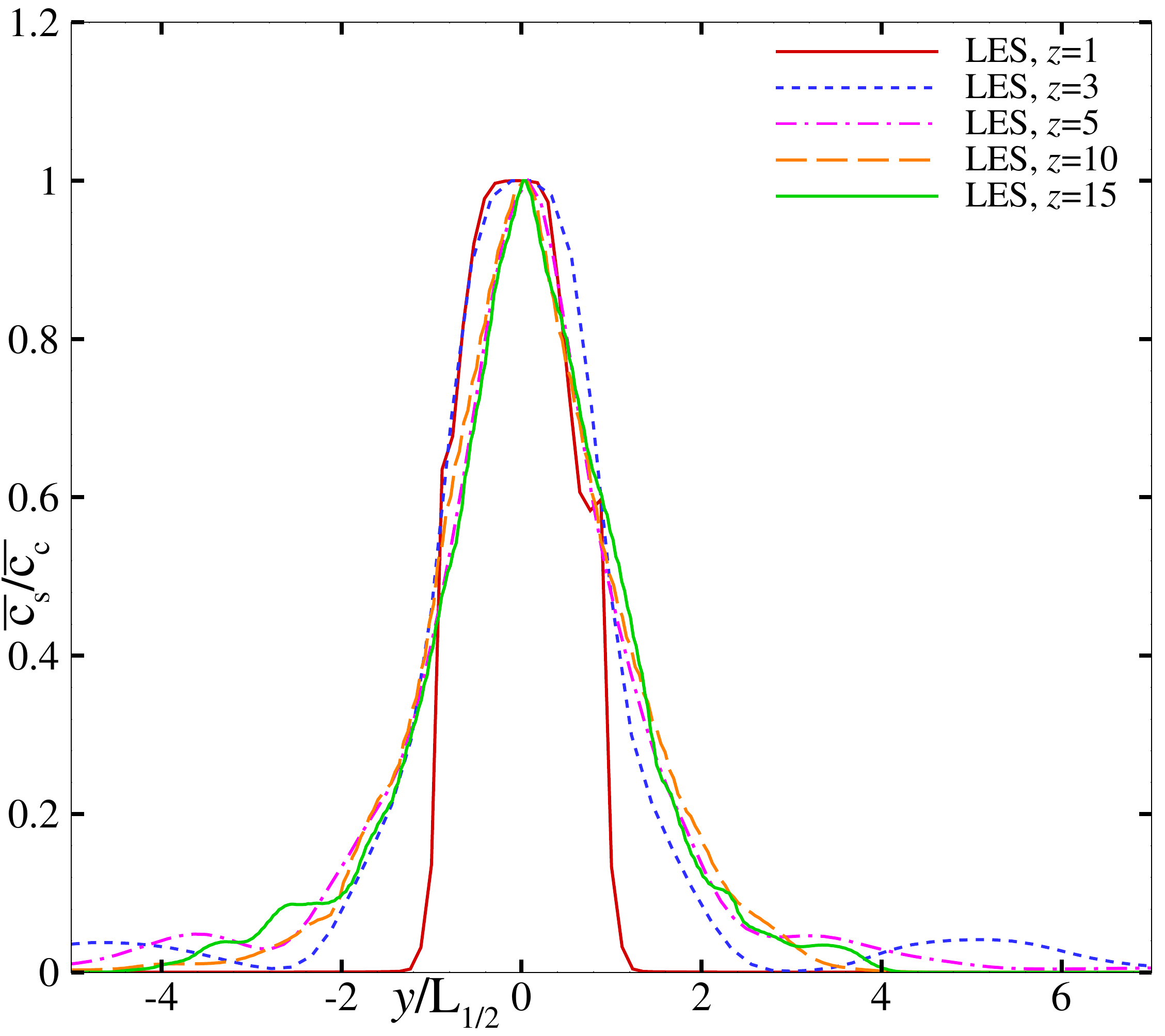}\\
	\caption{{Time-averaged} concentration profiles along jet centrelines, taken at various heights for air, helium, and hydrogen, {and} obtained from a) LES, OP \& 3D jet in $x$-$z$ plane and b) LES \& OP jet in $y$-$z$ planes.}
	\label{fig.Cs}
\end{figure}

\section{Discussion}
\subsection{Asymmetry of the jet}
\label{sec.asymmetry}
For the 3D jets, asymmetric flow structure was always observed. It was found that the perpendicular nature of the orifice, relative to the direction of flow within the pipe, caused a deflection of the jet away from the vertical $z$-axis.  It is not yet clear how the deflection angles scale for each gas. However, air was found to deflect more than helium (and hydrogen), despite having equal initial momentum flux (force) ejecting through the orifice.  Initially, from Fig.\ \ref{fig.centreline}, all three gases have very close deflection angles.  After only $z>2D$, was the experimental air jet found to deflect more than helium.  Upon considering the simulated trajectories, obtained from the C.M.\ locations, it was found that the heavier gases deflected more about the $z$-axis compared to lighter gases, with hydrogen having the least amount of deflection.  {Although other factors might contribute, the higher deflection of the air jet compared to the less dense helium and hydrogen jets is consistent with the relative strength of the corresponding vertical buoyancy forces.}

From the numerical simulations, in Fig.\ \ref{fig.centreline}, one notable `event' was always observed to occur for each gas, in the near field, between $z=5D$ to 10$D$.  Not only does this location correspond to the extent of the potential-core, in each case, but the $\lvert \boldsymbol{\overline{u}} \rvert_{\textrm{max}}$ location was found to switch sides relative to the C.M.  Up until $z\sim9D$ for air, and $z\sim6D$ for helium and hydrogen, the $\lvert \boldsymbol{\overline{u}} \rvert_{\textrm{max}}(z)$ were located to the right of the C.M.  At this point, however, the locations of $\lvert \boldsymbol{\overline{u}} \rvert_{\textrm{max}}$ remained almost constant in $x$ until $z\sim11D$ for all three gases.  Downstream, the {trajectories} of $\lvert \boldsymbol{\overline{u}} \rvert_{\textrm{max}}$ deflected to the right again, {but remained misaligned to the left of the C.M.}

\begin{figure}
	\centering
	\vspace*{10pt}
	\includegraphics[scale=0.4,trim={0.1cm 1.0cm 2.5cm 1.0cm},clip]{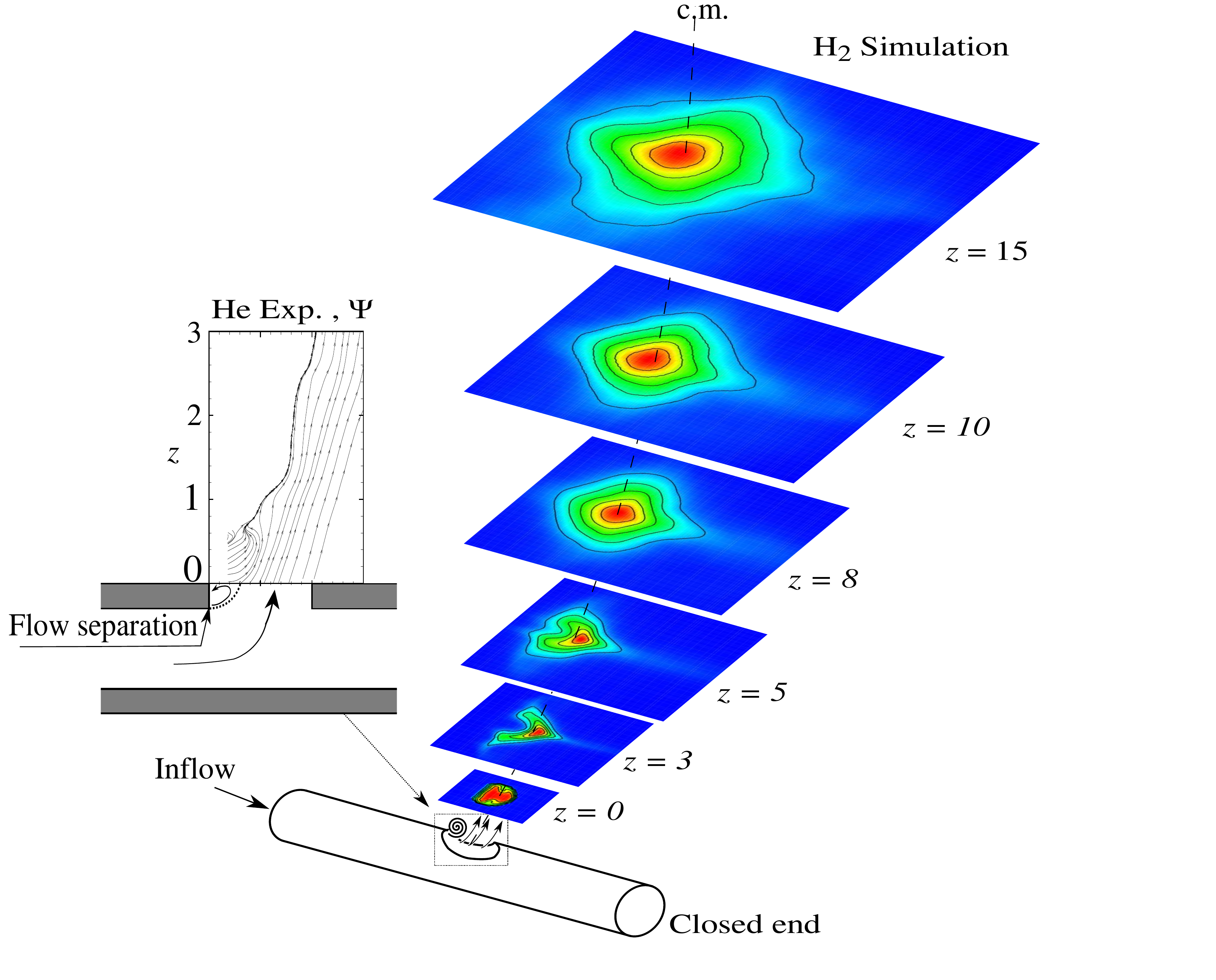}
	\caption{{Time-averaged} velocity contours in $x$-$y$ planes for the simulated $\textrm{H}_2$ jet. Also shown are instantaneous velocity streamlines ($\Psi$) obtained for the helium 3D jet, near the orifice.}
	\label{fig.U_profiles_z}
\end{figure}

In order to gain insight {into the reasons for shifts of} the location of $\lvert \boldsymbol{\overline{u}} \rvert_{\textrm{max}}(z)$ {with respect to the C.M.,} average velocity contours were extracted from the simulations on $x$-$y$ planes at various heights, along $z$, for all three gases.  Figure \ref{fig.U_profiles_z} shows these velocity contours for the hydrogen simulation, although it is noted that the other gases were qualitatively similar through their evolution.  At the very start of the jet evolution, shown at $z=0$, the initial jet was not circular.  In fact, a velocity deficit existed near the left-most portion of the orifice, and also {in} two regions on the right side, near $x=0.25D$ and $y\pm0.3D$.  The velocity deficit on the left of the jet likely resulted from flow separation of the moving gas originating from inside the pipe, akin to flow over a backward step \cite{Vogel1985JoHT922} or cavity \cite{Gharib1987JoFM501}. This phenomena was also observed in the 3D jet experiments and visualized in Fig.\ \ref{fig.U_profiles_z} with instantaneous velocity streamlines obtained from the helium experiment.  It {is} also likely that this flow separation encouraged entrainment of air on the back side of the jet (in the $+x$-direction) which lead to the enhanced mixing observed in this region.  For the other two flow deficit regions, near near $x=0.25D$ and $y\pm0.3$, these were likely caused by the curvature of the pipe diameter relative to the hole size; {further investigation is required to ascertain this.}

Downstream, at $z=3D$ in Fig.\ \ref{fig.U_profiles_z}, a {nearly stagnant region was formed} and centred at $x=y=0$.  Also, there existed {regions of} significant flow velocity {magnitudes on both sides of the stagnant region} in the $\pm y$ directions.   This explains why the C.M.\ of the jet was initially misaligned with the $\lvert \boldsymbol{\overline{u}} \rvert_{\textrm{max}}(z)$ location.  This flow pattern also explains the saddle-back {feature} observed in the velocity profiles of the experimental 3D jets, in the $y$-$z$ planes of Fig.\ref{fig.Velocity_Concentartion_Contours}.  The well-known Vena Contracta effect, generated immediately downstream from the orifice, may have also contributed to this saddle-back profile due to inward radial velocities at the edge of the jet\cite{Mi2001JoFE878}. Eventually, at $z=5D$, the {stagnant region became entrained} by the jet as mixing {occurred}.

By $z=8D$, although significant mixing and jet spreading had occurred by this point, a portion of the simulated jets remained asymmetric.  In fact, the overall shape of the simulated jets were stretched in the $+x$ direction, relative to the jet centre.  From $x>3D$, there appeared to {exist} minor secondary jetting along the $+x$ direction, which was not observed in any other directions.  This secondary jetting was believed to contribute to the far field misalignment of the $\lvert \boldsymbol{\overline{u}} \rvert_{\textrm{max}}(z)$ location with the C.M.  {However,} this secondary jetting was not observed in the {experiments,} likely due to fine-scale mixing near the orifice, which was not captured in the simulations.  Thus, it is very likely that the true C.M.\ of the jet {is infact} aligned with the $\lvert \boldsymbol{\overline{u}} \rvert_{\textrm{max}}(z)$ location in the far field.  

{At} $z=15D$, the shapes of the simulated helium and air jets had become elliptical (not shown).  {Significant} jet spreading was observed in the $+x$ direction compared to all other directions, {which is consistent with} the increased jet spreading along the $n$ direction relative to the $y$ direction,  as observed in Fig.\ \ref{fig.centrelineVelocity}b.  The hydrogen simulation, on the other hand, was found to develop into a fairly round jet by $z=15D$, as shown.  Although the {location of} $\lvert \boldsymbol{\overline{u}} \rvert_{\textrm{max}}(z)$ was still slightly misaligned with the C.M., the spreading rates in all directions were nearly equal.  Whether the air and helium jets eventually become axisymmetric, in the far field, remains to be investigated.  {The degree of asymmetric jet spreading observed for heavier gases, in the $n$ and $y$ directions, is consistent with buoyancy effects.  It is very likely that increased horizontal deflection associated with lower buoyancy forces leads to more interaction of the jet with the pipe boundary, thus contributing to the observed asymmetry.  It is also} possible that the early symmetric development for hydrogen, compared to air and helium, may arise due to enhanced mixing associated with the supersonic nature of the jet {in the former case}.

\subsection{Implications of jet asymmetry on ignition limits}

In the experimental concentration fields presented in Fig.\ref{fig.Velocity_Concentartion_Contours}, helium was found to have much higher concentration levels in the far field compared to air, {at} $z>3D$.  This {observation can be} attributed to {a} low Schmidt number ($Sc<1$), {in which case} mass diffusion rates are faster than momentum diffusion rates. On the other hand, axisymmetric OP jets exhibited even higher concentration levels, compared to 3D jets, in both near and far fields for both gases. This {result} is further {supports} the fact that {significantly} higher turbulent mixing and entertainment rates occur in the 3D jets compared to axisymmetry OP jets, thus shortening the extents of where the ignition limits might lie along the jet centreline in the far field, in the ($+s$-direction).  This {effect} is also evident {in} the faster velocity decay rates, and shortened potential core regions, as observed previously in Fig.\ \ref{fig.centrelineVelocity}.  While the the 3D jets ultimately developed into self-similar concentration profiles, as observed in Fig.\ \ref{fig.Cs}, such self-similarity cannot be used to predict the ignition limits in the near field.  This {implication} is especially true for the back side of the jets (in the $-x$-direction), for $z<5D$, where this enhanced mixing gave rise to higher concentration levels beyond $(n/L_{c,1/2}) < -1$.  Moreover, due to unsteady nature of the jet,  as observed from instantaneous concentration fields of Fig.\ref{fig.Instantaneous_plots}, there exists the possibility of increased local gas concentration, above the flammability-limit, in regions beyond those predicted by the self-similar profiles of Fig.\ \ref{fig.Cs}.  {A similar trend was} previously shown to be the case for axisymmetric jets \cite{Chernyavsky2011IJoHE2645}, where the {probability} of hydrogen ignition outside of conventional self-similar time-averaged limits was computed through simulation.  In future work, this {effect} should also be investigated for realistic jet configurations.

\subsection{Departures of simulation from experiment}

In this investigation, several discrepancies were observed between the simulations and experiments.  First, the appearance of {stagnant regions} in the {computed} flow patterns at $z=3D$ of Fig.\ref{fig.U_profiles_z} were not so prevalent in the experiments.  It was found that the experiments {exhibited more substantial} mixing, from the onset of release, compared to the simulations.  This {enhanced mixing} had the effect of mitigating the numerically observed {stagnant regions}, and also shortened the potential core length compared to the simulations.  It is well known that increases in turbulent mixing rates can reduce the potential core length of a jet \cite{Zaman1998PoF2652}.  Thus, it is probably the case that persisting laminar conditions exist in the potential core due to insufficient near field resolution in the simulations.    Despite this short-coming, the simulations {were} found to capture the correct trends observed experimentally and have provided some insight, physically, in terms of the asymmetric nature of the jet, which emerged radially from {the} pipe.

In terms of the velocity decay, the simulation captured well the experimental measurements for air.  However, a significant deviation from {the} experimental measurements was observed {in the case of} the helium simulation.  It is unclear why this departure between experiment and simulation existed.  It is possible that errors associated with mixing in the two-$\gamma$ model generated a faster velocity decay for helium compared to {the experimental case.}

Finally, it has been observed that the location of $\lvert \boldsymbol{\overline{u}} \rvert_{\textrm{max}}(z)$ was inconsistent between the simulations and {the} experiments. From {the} experiments, it was found that this location was relatively centred in the jet in the far field.  In the simulations, however, the location of $\lvert \boldsymbol{\overline{u}} \rvert_{\textrm{max}}(z)$ had a {tendency} to shift toward the left side of the jet C.M. location (in the $-x$ direction).  Despite this, the numerical prediction of the C.M. was found to agree well with the $\lvert \boldsymbol{\overline{u}} \rvert_{\textrm{max}}(z)$ location obtained from {the} experiments.   It is possible that longer sampling periods are required in order to accurately predict the position of $\lvert \boldsymbol{\overline{u}} \rvert_{\textrm{max}}(z)$ numerically.

\section{Conclusions}

{In this study}, experiments were conducted in order to investigate compressible turbulent jets, of varying gas densities and Reynolds numbers, issuing from realistic pipe geometry, and {to compare them} to axisymmetric round jets. A large-eddy-simulation strategy was also developed to provide further insight into the experimentally observed trends and {the} evolution {of the flow patterns} of the realistic 3D jets.  The fluids considered were air and helium for the experiments, and {the} simulations provided further insight into the behaviour of hydrogen.

It was found that {the} flow within a pipe, perpendicular to an upward facing hole, caused the resulting jet {outside the pipe} to form at a deflection angle relative to the vertical axis, in the direction of the {flow within the} pipe itself.  This deflection was influenced by {the} buoyancy of the jet, where heavier gases were found to deflect more than {the} lighter gases.  Furthermore, flow separation inside the pipe, at the orifice, and curvature of the pipe, relative to the size of the hole, {have contributed to} the asymmetric flow patterns observed.  In general, both air and helium {experienced significantly} more jet spreading compared to the axisymmetric jet experiments. Also, more jet spreading was observed on the back side of the asymmetric 3D jets compared to the axisymmetric case, in the near field.  This enhanced mixing in the asymmetric case caused a reduction in the potential-core length, and an increase in the velocity decay rate.   Also, in the far field, air and helium simulations were found to have {substantially} more jet spreading along the direction of the pipe, compared to all other directions.  Hydrogen, however, was found to {spread in a quasi-axisymmetric manner} in the far field.  Thus, caution is required when using round jet assumptions to describe the correct dispersion and velocity decay of a jet emitted from realistic geometries.

The authors would like to acknowledge the Natural Sciences and Engineering Research Council of Canada (NSERC) for {providing} project funding, and Compute Canada for {providing} computational resources.

\section*{References}

\bibliography{Literature_Review}

\end{document}